\documentclass[12pt]{article}
\usepackage{graphicx,psfrag,epsf}
\usepackage{enumerate}
\usepackage{url} 
\usepackage[english]{babel}
\usepackage[utf8x]{inputenc}
\usepackage[T1]{fontenc}
\usepackage{listings}
\usepackage{float}
\usepackage{natbib}
\usepackage{placeins}
\usepackage{amssymb}
\usepackage{xfrac}
\usepackage{enumerate, amsmath, graphicx, subfigure, rotating, verbatim,multicol}
\usepackage[ usenames,,dvipsnames ]{color}
\usepackage{float}

\usepackage{titling}

\usepackage{color}

\usepackage[margin=1in]{geometry}

\usepackage{algorithm}
\usepackage[noend]{algpseudocode}

\usepackage{mathtools}

\newcommand{\pkg}[1]{{\fontseries{b}\selectfont #1}} 

\makeatletter
\def\BState{\State\hskip-\ALG@thistlm}
\makeatother

\newcommand{\blind}{1}

\usepackage{times}

\usepackage{amsmath}
\usepackage[mathlines]{lineno}
\usepackage{etoolbox}

\newcommand*\linenomathpatchAMS[1]{
  \expandafter\pretocmd\csname #1\endcsname {\linenomathAMS}{}{}
  \expandafter\pretocmd\csname #1*\endcsname{\linenomathAMS}{}{}
  \expandafter\apptocmd\csname end#1\endcsname {\endlinenomath}{}{}
  \expandafter\apptocmd\csname end#1*\endcsname{\endlinenomath}{}{}
}

\expandafter\ifx\linenomath\linenomathWithnumbers
  \let\linenomathAMS\linenomathWithnumbers
  \patchcmd\linenomathAMS{\advance\postdisplaypenalty\linenopenalty}{}{}{}
\else
  \let\linenomathAMS\linenomathNonumbers
\fi

\linenomathpatchAMS{gather}
\linenomathpatchAMS{multline}
\linenomathpatchAMS{align}
\linenomathpatchAMS{alignat}
\linenomathpatchAMS{flalign}

\usepackage{authblk}
\usepackage{blindtext}

\begin{document}

\def\spacingset#1{\renewcommand{\baselinestretch}%
{#1}\small\normalsize} \spacingset{2}

\if1\blind
{
  \title{\bf{A Lattice and Random Intermediate Point Sampling Design for Animal Movement}}
  \author[1]{Elizabeth Eisenhauer}
\author[1]{Ephraim Hanks}
\affil[1]{Department of Statistics, Pennsylvania State University}
\affil[ ]{\textit {eisenhauer@psu.edu, hanks@psu.edu}}
 \maketitle
} \fi

\if0\blind
{
  \bigskip
  \bigskip
  \bigskip
  \begin{center}
    {\LARGE\bf Paper Review}
  \end{center}
  \medskip
} \fi

\begin{abstract}
    Animal movement studies have become ubiquitous in animal ecology for estimation of space use and analysis of movement behavior. In these studies, animal movement data are primarily collected at regular time intervals. We propose an irregular sampling design which could lead to greater efficiency and information gain in animal movement studies. Our novel sampling design, called lattice and random intermediate point (LARI), combines samples at regular and random time intervals. We compare the LARI sampling design to regular sampling designs in an example with common black carpenter ant location data, an example with guppy location data, and a simulation study of movement with a point of attraction. We modify a general stochastic differential equation model to allow for irregular time intervals and use this framework to compare sampling designs. When parameters are estimated reasonably well, regular sampling results in greater precision and accuracy in prediction of missing data. However, in each of the data and simulation examples explored in this paper, LARI sampling results in more accurate and precise parameter estimation, and thus better prediction of missing data as well. This result suggests that researchers might gain greater insight into underlying animal movement processes by choosing LARI sampling over regular sampling.

\end{abstract}

\section{Introduction}\label{sec:int}

Animal movement studies advance scientific knowledge of animal behavior in space and time. Insight from animal movement models helps researchers understand how animals interact with human and environmental factors. For example, researchers have conducted analyses of wildlife telemetry data to predict the affect of climate change on species range \citep{schloss2012} and to assess the impact of roadways on gene flow in terrestrial vertebrate populations \citep{shepard2008}. Furthermore, understanding the relationships between animals and their surroundings can benefit conservation efforts \citep{festa2003, chester2012conservation, berger2004last} and provide insight into disease dynamics \citep{wijeyakulasuriya2019extreme, conner2004movement}.

Researchers often record wildlife telemetry data at regular intervals \citep{weimerskirch2002gps, forester2007state, kareiva1983analyzing, parlin2018, roeleke2018, mcduie2019} and occasionally at higher frequencies at times when finer movement behavior is expected \citep{richardson2018}. There is evidence that increasing the frequency of regular samples greatly improves estimates of movement distance and territory size \citep{mills2006}, but resource limitations often lead to difficulties in consistently obtaining samples at high frequencies without reducing the overall length of the study. In this work, we show that the use of sampling designs other than regular sampling can lead to better inference on parameters in animal movement models without requiring additional samples or reducing study duration. While \citet{millspaugh2001radio} mention the application of a range of sampling designs with stochastic components for wildlife telemetry studies, these designs have not been thoroughly compared in context and are rarely implemented.

There is a widely accepted view in geostatistics that samples at regular intervals in space lead to better interpolation of data, while clustered samples lead to better estimation of spatial covariance parameters \citep{zimmerman}. To compromise between parameter estimation and prediction at unobserved locations, \citet{zimmerman} suggests inclusion of samples with regular spacing as well as groups or pairs of points that are close together. One example of this is a lattice plus close pairs approach in which at least half of the locations form a regular lattice in 2D space, while the remaining points are randomly assigned within a disc centered at randomly selected lattice locations \citep{diggle}. \citet{diggle} found that a lattice plus close pairs design performed better than a lattice alone or a lattice and infill approach in regard to spatial prediction. 

Theoretical support for sampling at different scales is lacking in the geostatistical literature. On the contrary, systematic or regular sampling has been shown to be optimal within some subclasses of two-dimensional sampling designs in regard to minimizing variance of the sample mean \citep{bellhouse1977some}. However, few problems remain in animal movement modeling in which the only goal is precise estimation of a population mean. Instead, we look to the experimental literature which evidences the superiority of irregular sampling for detection of spatial patterns \citep{fortin1990spatial, oliver1986combining}.

In this paper, we propose a sampling scheme for animal telemetry data inspired by the lattice plus close pairs geostatistical design. Our proposed approach, which we call a lattice and random intermediate point (LARI) design, requires data collection at regular time intervals coupled with one randomly selected time point in between each adjacent pair of regular samples. We conjecture that the regular time intervals will result in suitable temporal coverage while the random intermediate points will capture behavior at short time lags. We suspect that capturing behavior at different time scales will correspond with improved estimation of movement parameters.

This LARI sampling design was motivated by a problem that arose in collection of wood nesting carpenter ant, $Camponotus$ $pennsylvanicus$, movement data at the Pennsylvania State University. Members of the Hughes laboratory captured video footage of ants in a wooden nest over a 4 hour time frame and recorded coordinate locations of the ants at 1 second intervals \citep{modlmeier2019ant}. The data collection procedure was manually expensive, requiring the recruitment, training, and labor of seventeen undergraduate students \citep{modlmeier2019ant}. As a new experiment was planned involving a large number of nests over a longer time frame, it became apparent that the data collection strategy previously employed would not be feasible at the necessary scale. Thus we set out to develop a sampling design that would result in similar model inference while reducing the manual cost. Of course, this motivation is not limited to the ant example. Restrictions on data collection frequency and magnitude are commonplace in animal movement studies, especially those that employ tracking devices \citep{tomkiewicz2010global}.

We describe the LARI sampling scheme in detail in Section \ref{sec:sam}. In Section \ref{sec:sto} we outline a stochastic differential equation (SDE) model for movement similar to that of \citet{russell2018}. In Section \ref{sec:sim}, we compare parameter estimation and prediction accuracy between sampling designs via a simulated example. In Section \ref{sec:gup}, we compare parameter estimates between sampling designs using subsamples of guppy movement data. In Section \ref{sec:com}, we present a novel modeling framework which we apply to the high resolution carpenter ant movement data and implement to compare sampling designs.

\section{A Lattice and Random Intermediate Points Sampling Scheme}\label{sec:sam}

In a given animal movement study, assume data collection is set to begin at time $0$ and end at time $T$. Assume resources are limited and only $n$ samples will be collected in this time frame for a single individual. Sampling the animal's position at regular time intervals of length $h = \frac{T}{n-1}$ results in the data matrix
\begin{equation}
\mathbf{D}_{\text{Regular}}\equiv\begin{bmatrix}
   0 & h & 2h & \hdots & T-h & T \\
   \mathbf{r}_0 & \mathbf{r}_h & \mathbf{r}_{2h} & \hdots & \mathbf{r}_{T-h} & \mathbf{r}_T
    \end{bmatrix}' \label{eq:sam010}    
\end{equation}
where 
$\mathbf{r}_{t} \equiv \begin{bmatrix}
    x_{t} & y_{t}
\end{bmatrix}'$ 
is the $x$- and $y$-coordinate vector of the animal's position at time $t \in \{0, h, 2h, \hdots, T-h, T\}$. While regular sampling minimizes the maximum time between observations, movement behavior occurring at finer time scales than those sampled is not captured in the observed data.

We propose a lattice and random intermediate points (LARI) sampling scheme, which produces the data matrix
\begin{equation}
\mathbf{D}_\text{LARI}\equiv\begin{bmatrix}
    0 & t_0^* & 2h & t_1^* & 4h & \hdots & T-2h & t_{\frac{n}{2}}^* & T \\
    \mathbf{r}_0 & \mathbf{r}_{t_0^*} & \mathbf{r}_{2h} & \mathbf{r}_{t_1^*} & \mathbf{r}_{4h} & \hdots & \mathbf{r}_{T-2h} & \mathbf{r}_{t_{\frac{n}{2}}^*} & \mathbf{r}_T
    \end{bmatrix}'\label{eq:sam020}    
\end{equation}
where
    \begin{align*}
        t_i^*\sim \text{Uniform}\left(2 h i, 2 h (i+1)\right), \quad i \in \left\{0, 1, 2, \hdots, \frac{n}{2} \right\}.
    \end{align*}
In practice, it may be more realistic to choose $t_i^*$ from a Discrete Uniform distribution depending on the sampling resolution. 

Both data matrices $\mathbf{D}_\text{Regular}$ and $\mathbf{D}_\text{LARI}$ contain $n$ observations for a single individual. To collect data for multiple individuals over multiple time frames, repeat this procedure as necessary.

\section{Stochastic Differential Equation Model for Animal Movement}\label{sec:sto}

We follow \citet{russell2018} and \citet{hanks2017} and consider a flexible stochastic differential equation (SDE) model for an animal's position $\mathbf{r}_t$ at time $t$
\begin{align}
    d\mathbf{r}_t&=\mathbf{v}_t dt\label{eq:sde010}\\
    d\mathbf{v}_t&=-\beta(\mathbf{v}_t-\boldsymbol{\mu}(\mathbf{r}_t))dt+c(\mathbf{r}_t)\mathbf{I}d\mathbf{w}_t\label{eq:sde020}
\end{align}
where $\mathbf{v}_t$ is the velocity of the animal at time $t$, $\beta$ is the coefficient of friction \citep{nelson1967} which controls autocorrelation in movement, $\boldsymbol{\mu}(\mathbf{r}_t)$ is the mean drift in the direction of movement, $c(\mathbf{r}_t)$ is a scalar that controls the variance in the stochastic term, $\mathbf{I}$ is a $2\times 2$ identity matrix, and $\mathbf{w}_t$ is independent Brownian motion in $\mathbb{R}^2$. This SDE framework is attractive because of the wide range of movement behavior which can be modeled. For example, the right hand side of \eqref{eq:sde020} can be viewed as the sum of forces acting on the animal at time $t$ and at position $\mathbf{r}_t$. For instance, there could be a force toward the center of the animal's home range, toward the nearest food source, toward breeding grounds, toward higher or lower elevation, away from the nearest predator, or away from cooler temperatures. Depending on the time frame and study species, these forces could vary over time or space.

\begin{figure}[h]
\centering\includegraphics[width=0.4\linewidth]{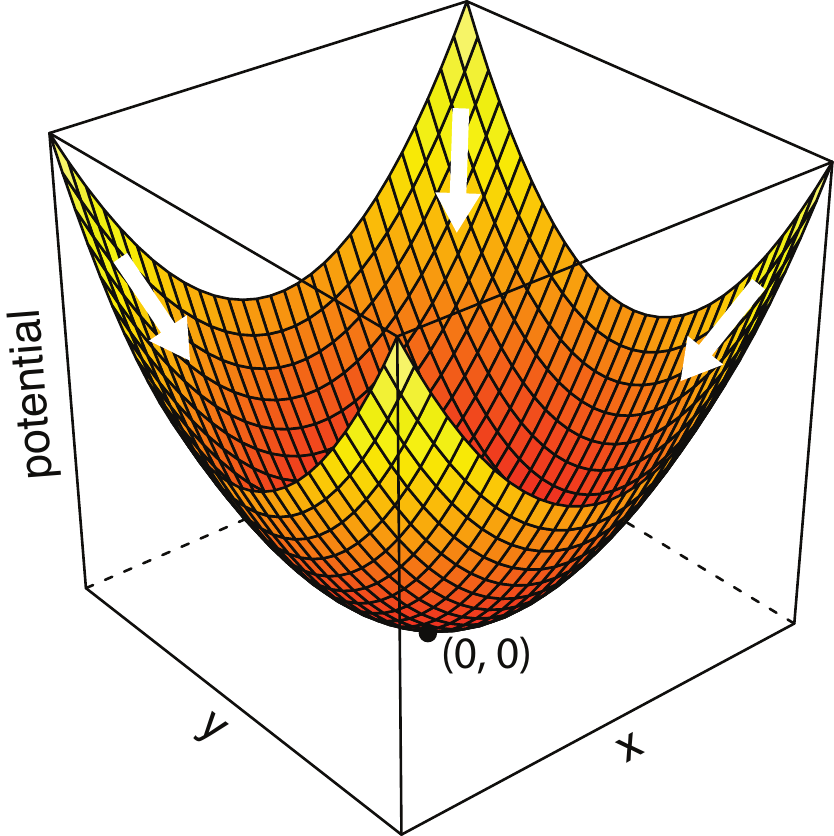}
\caption{Quadratic potential surface with a single attraction point at the origin.}
\label{fig:quad_pot}
\end{figure}

\citet{brillinger2001a} used a similar SDE framework and adopted potential functions from particle and planetary movement models to model elk movement by setting $\boldsymbol{\mu}(\mathbf{r}_t) = -\triangledown p(\mathbf{r}_t)$, the negative gradient of a potential surface $p(\mathbf{r}_t)$. The potential surface is a continuous surface or grid with the highest values on the surface at repulsive locations, lowest values at attractive locations, and relatively central values in areas where the force is neutral. Under this model, the average animal in a population moves around the space avoiding those points of repulsion or areas with high potential and moving toward points of attraction or areas with low potential. A simple example of a potential surface is the quadratic function $k(x^2+y^2)$ which will be used in a simulation example in Section \ref{sec:sim}. This quadratic potential surface has a single point of attraction at the origin, as shown in Figure \ref{fig:quad_pot} with the parameter $k = 1$. The white arrows displayed in Figure \ref{fig:quad_pot} point down the gradient of the potential surface, in the direction of mean drift. One might utilize this potential surface in a model for movement of a central place forager, with movement centered around $\begin{bmatrix}     0&0 \end{bmatrix}'$ and $k$ controlling the strength of attraction to this central location. Potential surfaces can be much more complex than this example, as we will see in Section \ref{sec:com}. For further detail on the use of potential surfaces to model animal movement, see \citet{preisler2013analyzing}.

\citet{russell2017modeling, russell2018} expanded the SDE framework of \citet{brillinger2001a} to include motility surfaces, which describe overall speed independent of direction as a function of location. The motility surface is a surface or grid of values assigned on the space inhabited by the animal. High motility values are indicative of fast movement or high speed in the corresponding location. Low motility is indicative of slow movement.

The SDE model we define in this section is similar to that of \citet{russell2018} with zero measurement error and assuming the motility surface is smooth. As in \citet{russell2018}, we define the mean drift and magnitude of stochasticity with spatially-varying motility and potential surfaces. The potential surface $p(\mathbf{r}_t)$ captures spatially-varying directional bias (drift) through $-\triangledown p(\mathbf{r}_t)$, while the motility surface $m(\mathbf{r}_t)$ models spatial variation in speed without directional bias by compressing and dilating time. The mean drift $\boldsymbol{\mu}(\mathbf{r}_t)$ and magnitude of stochasticity $c(\mathbf{r}_t)$ are defined
\begin{align}
    \boldsymbol{\mu}(\mathbf{r}_t)&\equiv m(\mathbf{r}_t)[-\triangledown p(\mathbf{r}_t)]\label{eq:sde030}\\
    c(\mathbf{r}_t)&\equiv \sigma m(\mathbf{r}_t)\label{eq:sde040}
\end{align}
where $\sigma$ controls the magnitude of the random forces acting on the animal.
We chose to ignore measurement error because the measurement error in our ant data is negligible. As sophisticated technology allows for greater accuracy in animal tracking, we expect the need for measurement error specification for animal location to diminish. When movement error is not negligible, state-space models can be used with the SDE model \eqref{eq:sde010}--\eqref{eq:sde040} being a model for the true, but latent, animal position over time.

\subsection{Numerical Approximations}

A closed-form solution to \eqref{eq:sde010}--\eqref{eq:sde040} only exists for very simple choices of $m(\cdot)$ and $p(\cdot)$. There is no closed-form solution whenever spatial constraints are present (e.g., \citet{hanks2017, russell2018}). In this section, we present a general numerical approximation to the SDE which is applicable in a broad range of settings including those where there is no closed-form solution.

\citet{hanks2017} and \citet{russell2018, russell2017modeling} describe numerical approximations using samples at regular time intervals and do not consider irregular time lags between samples. Our discrete-time approximation approach is similar to that of \citet{russell2018}, but we extend their framework to the case where the intervals between observation times can vary. Developing numerical methods for irregular time intervals will make inference more straightforward when data are missing or irregularly sampled. To simplify notation for irregular samples, we now change the subscript in equations from continuous time $t$ to ordered observation number $\tau$. Henceforth, $\mathbf{r}_\tau$ is the vector of elements in column $\tau$, row 2 of a data matrix of the form \eqref{eq:sam010} or \eqref{eq:sam020}.

Euler-Maruyama approximations are derived from Taylor series expansions \citep{kloeden2013numerical} and are commonly used to numerically approximate SDE models because they are easy to calculate. The Euler-Maruyama method approximates \eqref{eq:sde010}--\eqref{eq:sde020} by
\begin{align}
    \mathbf{r}_{\tau+1}&=\mathbf{r}_\tau +\mathbf{v}_\tau h_\tau\label{eq:num010}\\
    \mathbf{v}_{\tau+1}&=\mathbf{v}_\tau-\beta(\mathbf{v}_\tau-\boldsymbol{\mu}(\mathbf{r}_\tau))h_\tau+c(\mathbf{r}_\tau) \mathbf{I}d\mathbf{w}_\tau\label{eq:num020}
\end{align}
where $h_\tau$ is the change in time from observation $\tau$ to observation $\tau+1$. Here our approach differs from the SDE model of \citet{russell2018} where $h_\tau$ was constant with respect to $\tau$. Substituting \eqref{eq:num010} into \eqref{eq:num020} following \citet{hanks2017} results in
\begin{align}
{\frac{\mathbf{r}_{\tau+2}-\mathbf{r}_{\tau+1}}{h_{\tau+1}}-\frac{\mathbf{r}_{\tau+1}-\mathbf{r}_{\tau}}{h_\tau}}&=\beta h_\tau \left(\boldsymbol{\mu}(\mathbf{r}_\tau)-\frac{\mathbf{r}_{\tau+1}-\mathbf{r}_\tau}{h_\tau}\right)+c(\mathbf{r}_\tau) h_\tau^{\sfrac{1}{2}} \boldsymbol{\epsilon}_\tau \label{eq:num030}
\end{align}
where $\boldsymbol{\epsilon}_\tau\overset{\text{iid}}{\sim}N(\mathbf{0},\mathbf{I})$ and $\mathbf{0}$ is the zero vector in $\mathbb{R}^2$. This can be re-expressed as
\begin{align}
    \mathbf{r}_{\tau+2} &= \mathbf{r}_{\tau+1} + \frac{h_{\tau+1}}{h_\tau} \left( \mathbf{r}_{\tau+1} - \mathbf{r}_\tau \right) + \beta h_\tau h_{\tau+1} \left( \boldsymbol{\mu}(\mathbf{r}_\tau) - \frac{\mathbf{r}_{\tau+1} - \mathbf{r}_\tau}{h_\tau} \right) + c(\mathbf{r}_\tau)h_\tau^{\sfrac{1}{2}}h_{\tau+1}\boldsymbol{\epsilon}_\tau,\label{eq:10}
\end{align}
an equation in which the ant's position is a function of the two previous observed positions. 

In Supplemental Material \ref{sec:appendix:ex}, we provide examples of potential and motility surfaces, which we simulate from using \eqref{eq:10}. These examples illustrate how changing the motility and potential surfaces effects the movement behavior described by the model.

\section{Simulation Example}\label{sec:sim}

\subsection{Simulation from an SDE Model with Quadratic Potential Function}

We conducted a simulation example to compare the sampling schemes in \eqref{eq:sam010} and \eqref{eq:sam020}. We simulated data at a fine temporal scale from an SDE model with a quadratic potential function and constant motility surface. The quadratic function biases movement toward a single attraction point at $\begin{bmatrix}
    0&0
\end{bmatrix}'$. This approximates real movement behavior exhibited by central place foragers such as white-tailed deer \citep{tierson1985}. In this example,
\begin{align*}
    m(\mathbf{r}_\tau) &\equiv 1 \\
    p(\mathbf{r}_\tau) &\equiv k \mathbf{r}_\tau'\mathbf{r}_\tau 
\end{align*}
where $k\in \mathbb{R}$ controls the strength of attraction to the central location $\begin{bmatrix}
    0&0
\end{bmatrix}'$. Consequently,
\begin{align*}
    \boldsymbol{\mu}(\mathbf{r}_\tau) &= -\triangledown p(\mathbf{r}_\tau) = -2k\mathbf{r}_\tau\\
    c(\mathbf{r}_\tau) &= \sigma. 
\end{align*}
The set of SDEs \eqref{eq:sde010} and \eqref{eq:sde020} become
\begin{align}
    d\mathbf{r}_t&=\mathbf{v}_t dt\label{simquadsde1}\\
    d\mathbf{v}_t&=-\beta[\mathbf{v}_t-(-2k\mathbf{r}_\tau)] dt+\sigma\mathbf{I}d\mathbf{w}_t\label{simquadsde2}
    \end{align}
and the numerical approximation \eqref{eq:num030} becomes
\begin{align}
    {\frac{\mathbf{r}_{\tau+2}-\mathbf{r}_{\tau+1}}{h_{\tau+1}}-\frac{\mathbf{r}_{\tau+1}-\mathbf{r}_{\tau}}{h_\tau}}&=\beta h_\tau \left(-2k\mathbf{r}_\tau-\frac{\textbf{r}_{\tau+1}-\mathbf{r}_\tau}{h_\tau}\right)+h_\tau^{\sfrac{1}{2}}\sigma \boldsymbol{\epsilon}_\tau.\label{eq:sim050}
\end{align}
Since the simulated data is generated at regular time steps, we set $h_\tau = h$ for all observations $\tau$ and solve for $\mathbf{r}_{\tau+2}$ to get
\begin{align}
    \mathbf{r}_{\tau+2}&=\mathbf{r}_{\tau+1}(2-\beta h)+\mathbf{r}_\tau(\beta h-1-2\beta k h^2)+h^{\sfrac{3}{2}} \sigma \boldsymbol{\epsilon}_\tau \label{eq:sim060}
\end{align}
which is an autoregressive model of order 2. 

\begin{figure}[h]
\centering\includegraphics[width=\linewidth]{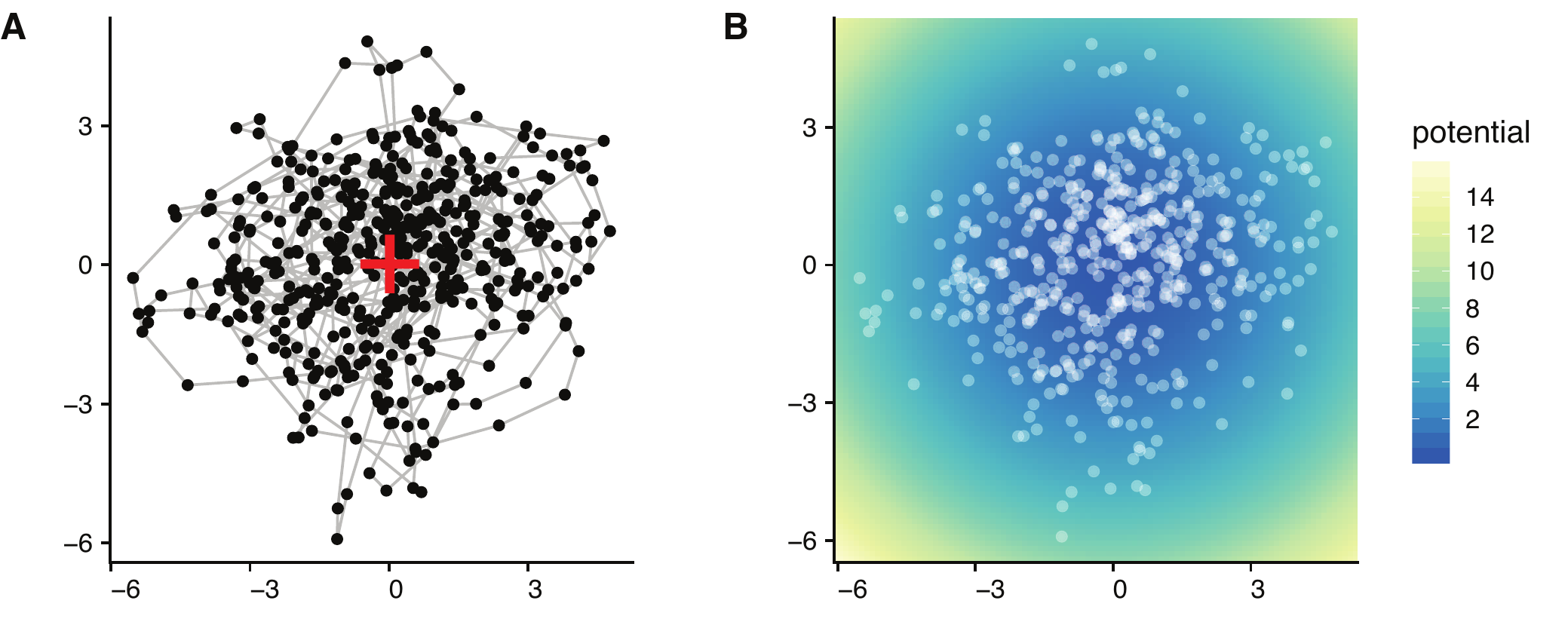}
\caption{(A) Simulated data for an individual with grey lines connecting those data points that are adjacent in time. The single attraction point is displayed as a red "+". (B) The quadratic potential surface with simulated positions for a single individual depicted in white.}
\label{fig:sim010}
\end{figure}

We simulated movement data for one individual over $n=500$ time points with time step $h=1$ and model parameters $\beta=0.4$, $\alpha\equiv k\beta=0.08$, and $\sigma=0.5$. Since simulation of observation $\tau$ requires observations $\tau - 1$ and $\tau - 2$ as input, we fixed the positions at the first two time points near the point of attraction $\begin{bmatrix}
    0&0
\end{bmatrix}'$. Specifically, $\mathbf{r}_1 = \mathbf{r}_2 = \begin{bmatrix}
    1&1
\end{bmatrix}'$. The next 498 time points were simulated recursively from \eqref{eq:sim060}. Figure \ref{fig:sim010} depicts one path simulated with this procedure. We will simulate 150 paths and consider each path separately. We will compare regular and LARI sampling schemes by subsampling each simulated path using \eqref{eq:sam010}--\eqref{eq:sam020} with $h = 5$ and comparing subsamples. 

We refer to the full simulation containing all $x$- and $y$-coordinates by $\{\mathbf{r}\}$. The general notation for the positions included in the LARI or regular subsample is $\{\mathbf{r}\}_\text{obs}$ and the positions removed by the subsampling procedure are $\{\mathbf{r}\}_{\text{unobs}}$. Thus, the subsampling procedure is represented by \[\{\mathbf{r}\}_\text{obs}=\{\mathbf{r}\}\setminus\{\mathbf{r}\}_{\text{unobs}}.\]

\subsection{Parameter Identifiability}\label{sec:theo}

In \citet{kloeden2013numerical}, the authors derive vector and matrix ordinary differential equations for the vector mean and second moment of a general vector linear SDE. In this subsection, we will interpret this derivation in the context of \eqref{simquadsde1} and \eqref{simquadsde2}. For simplicity, we describe this result in the $x$ direction only, where $r_x(t)$ is the $x$ component of the coordinate vector $\mathbf{r}_t$, $v_x(t)$ is the $x$ component of the coordinate vector $\mathbf{v}_t$, and $w_{x}(t)$ is independent Brownian motion in $\mathbb{R}^1$. 

We are interested in determining whether the parameters we intend to estimate ($\beta$, $\sigma$, and $\alpha$) are identifiable as we approach the stationary distribution (i.e., as $t\rightarrow \infty$). By combining \eqref{simquadsde1} and \eqref{simquadsde2}, we obtain the vector SDE
\begin{align}
d \begin{bmatrix} 
r_x(t)\\ 
v_x (t)
\end{bmatrix} &=
\begin{bmatrix}
0 & 1\\
-2\alpha & -\beta
\end{bmatrix}
\begin{bmatrix}
r_x\\
v_x
\end{bmatrix}_t dt +
\begin{bmatrix}
0\\
\sigma
\end{bmatrix} dw_{x}(t),
\end{align}
which allows us to derive the vector mean 
\begin{align}
\mathbf{n}(t) = \text{E}\left(\begin{bmatrix} 
r_x(t)\\ 
v_x (t)
\end{bmatrix}\right) = \begin{bmatrix}
0 & 1\\
-2\alpha & -\beta
\end{bmatrix}^{-1} \frac{d\mathbf{n}(t)}{dt}.
\end{align}
As we approach the stationary distribution and $\frac{d\mathbf{n}(t)}{dt} = 0$, the mean vector $$\mathbf{n}(t) = 0.$$ Therefore, the second moment of the stationary distribution
\begin{align}
S(t) = 
\text{E}\left(
\begin{bmatrix} 
r_x(t)\\ 
v_x (t)
\end{bmatrix}
\begin{bmatrix} 
r_x(t)\\ 
v_x (t)
\end{bmatrix}'
\right)
= \text{Var}\left(\begin{bmatrix} 
r_x(t)\\ 
v_x (t)
\end{bmatrix}\right)
\end{align}
is found by solving the system of equations
\begin{align}
\frac{dS(t)}{dt} = 
\begin{bmatrix}
0 & 1\\
-2\alpha & -\beta
\end{bmatrix}
S(t) + S(t)
\begin{bmatrix}
0 & 1\\
-2\alpha & -\beta
\end{bmatrix} +
\begin{bmatrix}
0\\
\sigma
\end{bmatrix}
\begin{bmatrix}
0\\
\sigma
\end{bmatrix}'.\label{theosyseq}
\end{align}
The stationarity of the distribution implies $\frac{dS(t)}{dt} = 0$, which along with \eqref{theosyseq} yeilds
\begin{align}
S(t) = \begin{bmatrix}
\frac{\sigma^2}{2\beta} & 0\\
0 & \frac{\sigma^2}{4\beta\alpha}
\end{bmatrix}.
\end{align}
Thus, we have 2 equations and 3 unknowns, rendering $\sigma$, $\beta$, and $\alpha$ unidentifiable.

This result highlights the value of having samples at short time lags. When telemetry data are sampled regularly at large time lags, the transient distribution will be well-approximated by the stationary distribution, and parameters in the model may become unidentifiable, or only weakly identifiable. Thus, we expect the regular subsample will lead to unidentifiability in parameter estimation. In Section \ref{sec:est}, we outline the model-fitting procedure applied to the LARI and regular subsamples. 

\subsection{Estimation of Model Parameters and Missing Values}\label{sec:est}

Initially, we attempted posterior approximation of the model parameters ignoring the missing data. However, this approach led to poor parameter inference (see Supporting Material \ref{sec:appendix:sim}), which could be due to the large reduction in movement variability that occurs when data are subsampled. These results led to our decision to estimate the positions at unobserved time points, thus reintroducing an appropriate amount of variability into the movement paths. 

To estimate $\beta$, $\alpha$, $\sigma$, and $\{\mathbf{r}\}_{\text{unobs}}$, we took a Bayesian approach and constructed an MCMC algorithm to sample from the joint posterior $\pi(\alpha,\beta,\sigma,\{\mathbf r\}_{\text{unobs}}|\{\mathbf r\}_\text{obs})$. For details on the posterior distribution and MCMC sampler, see Supporting Material \ref{sec:appendix:bay}.

\subsection{Simulation Example Results}

We simulated 150 paths, subsampled the paths using both a LARI and regular design, and individually fit each subsample using the estimation approach described in Section \ref{sec:est}. We assessed convergence of the MCMC algorithm in each case using the Geweke diagnostic \citep{geweke1991}. The Geweke convergence diagnostic quantifies the dissimilarity of the means of the first 10\% and last 50\% of the Markov chain iterations. In the Geweke diagnostic, the test statistic for variable $\eta$ is a $z$-score 
\[z=\frac{\bar{\eta}^{\text{(first 10\%)}}-\bar{\eta}^{\text{(last 50\%)}}}{\widehat{SE}}\]
where $\bar{\eta}^{\text{(first 10\%)}}$ is the sample mean of the first 10\% of the Markov chain, $\bar{\eta}^{\text{(last 50\%)}}$ is the sample mean of the last 50\% of the chain, and ${\widehat{SE}}$ is the asymptotic standard error of the difference, computed using spectral density estimates for the two sections of the chain. 

For each subsample and each simulated path, the Geweke diagnostic was computed for the three parameters $\alpha$, $\beta$, and $\sigma$. We labelled a subsample "converged" if the absolute values of the Geweke diagnostics for all three chains were less than 3. By this definition, of the 300 subsamples, 76.3\% converged. Specifically, 84\% of the LARI subsamples and 68.7\% of the regular subsamples converged. We removed all simulations where at least one subsample (regular or LARI) did not converge, leaving 87 of the 150 simulations for analysis. 

\begin{figure}[h]
\centering\includegraphics[width=.9\linewidth]{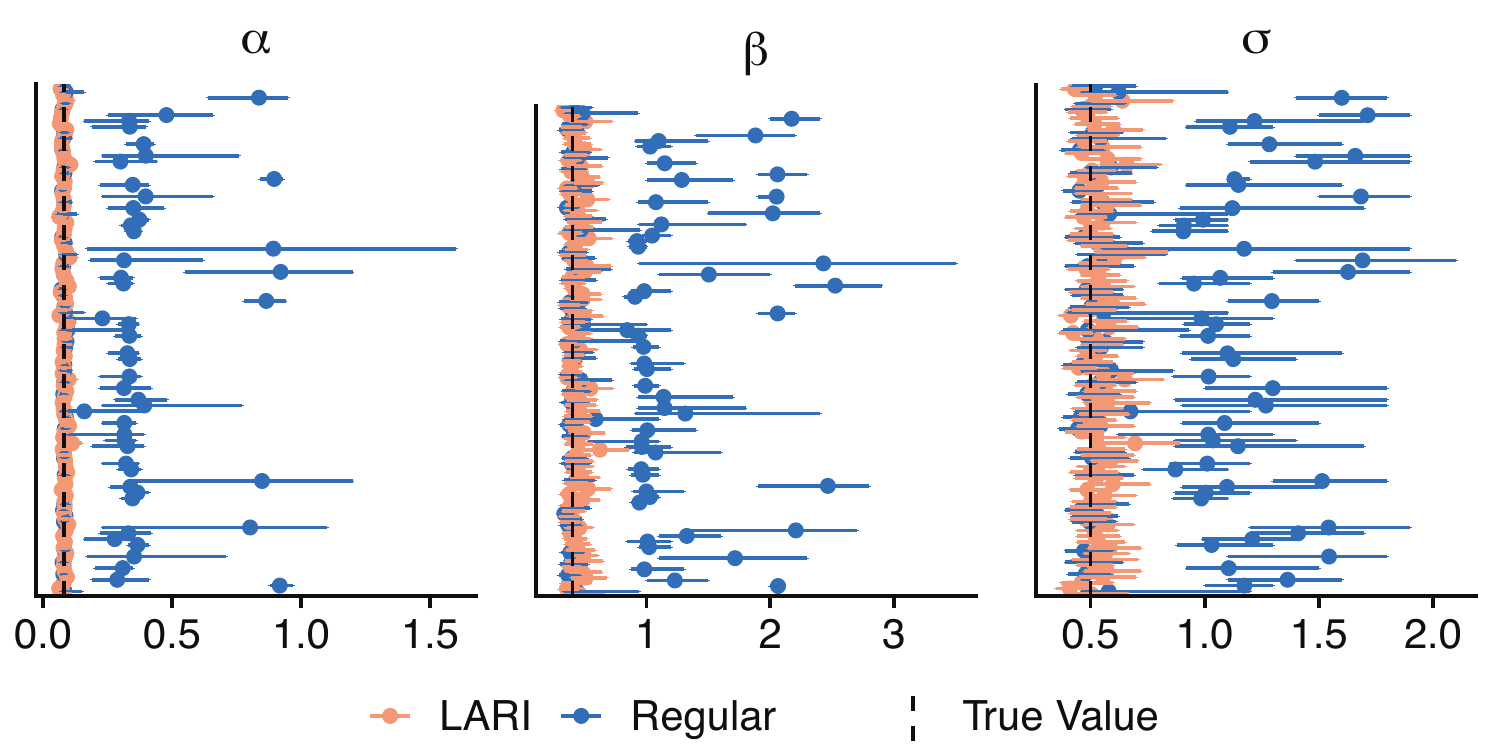}
\caption{95\% equal-tailed credible intervals for each subsample.}
\label{fig:converged_CIs}
\end{figure}

95\% equal-tailed credible intervals for the model parameters are shown for the 174 remaining subsamples in Figure \ref{fig:converged_CIs}. As depicted in Figure \ref{fig:converged_CIs}, many of the regular subsamples result in poor estimation of the model parameters. Only 46.6\% of the 87 regular subsamples resulted in credible intervals containing all three true parameter values, compared to 78.4\% of the 87 LARI subsamples. This result is consistent with the theoretical justification in Section \ref{sec:theo}, which suggests unidentifiability when we use a regular subsample at large enough time steps.

\begin{figure}[h]
\centering\includegraphics[width=.9\linewidth]{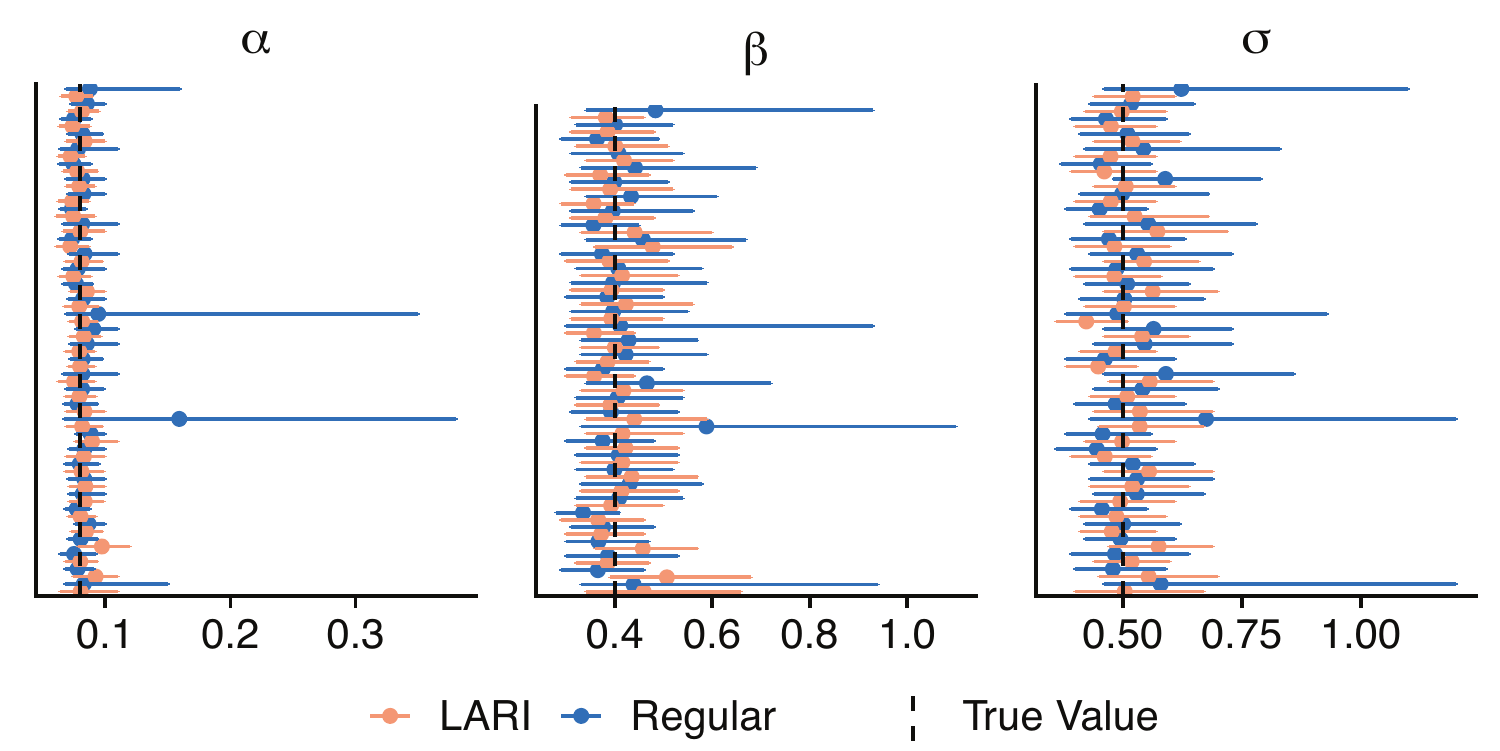}
\caption{95\% equal-tailed credible intervals for the "best case" subset.}
\label{fig:best_case_CIs}
\end{figure}

Although we have already obtained evidence in favor of LARI sampling for parameter estimation, we are also interested in the "best case" scenario where both subsamples capture the true parameter values in their 95\% credible intervals. There are 34 remaining simulations in this "best case" subset. The model parameter 95\% credible intervals for the "best case" subsamples are shown in Figure \ref{fig:best_case_CIs}.  

\begin{figure}[h]
\centering\includegraphics[width=.9\linewidth]{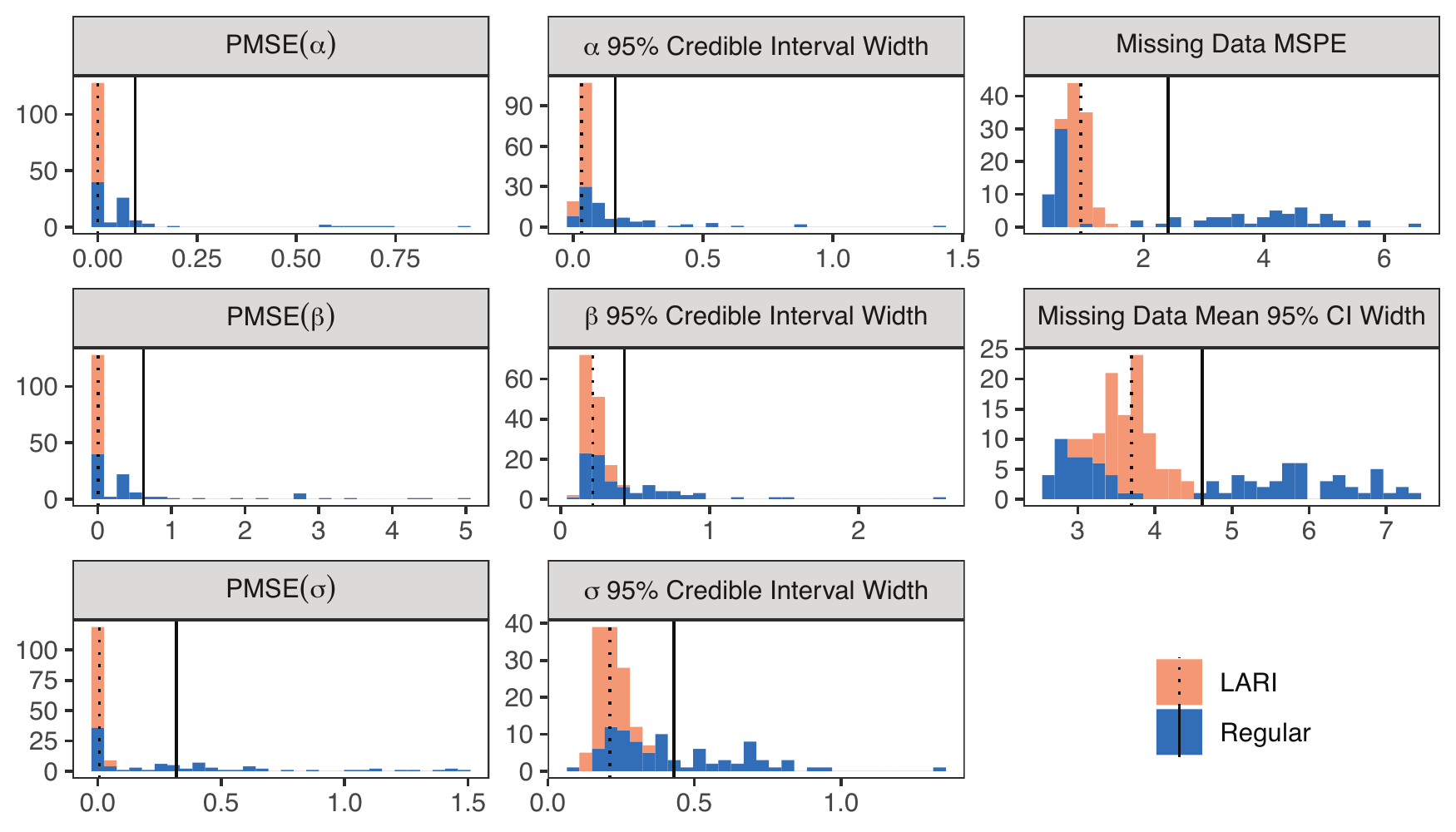}
\caption{Stacked histograms using all converged simulations (the absolute values of Geweke $Z$-scores for $\alpha$, $\beta$, and $\sigma^2$ were all less then 3). The dotted line delineates the mean of the LARI subsamples, and the solid line portrays the mean of the regular subsamples.}
\label{fig:res030}
\end{figure}

\begin{figure}
\centering\includegraphics[width=.9\linewidth]{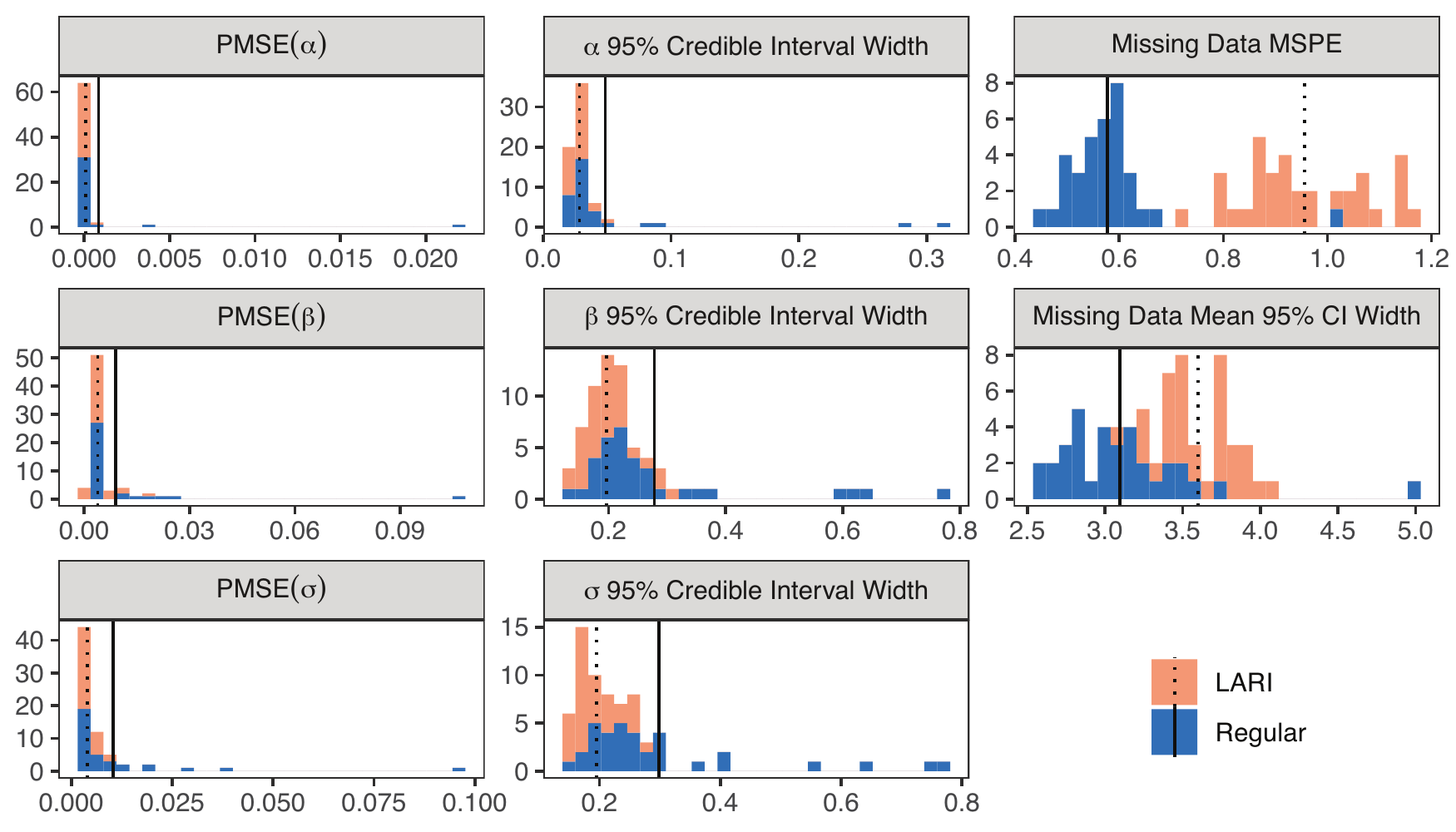}
\caption{Stacked histograms from the "best case" simulations. The dotted line is positioned at the mean of the LARI subsamples and the solid line at the mean of the regular subsamples.}
\label{fig:res040}
\end{figure}

We are now limited to 2 subsets; the subsamples which led to convergence by our definition and the "best case" subsamples. For each of these subsets, we will compare parameter estimation and prediction of missing values between the LARI and regular sampling designs with 8 metrics. We define these metrics in the following passages and display our results in Figures  \ref{fig:res030} and \ref{fig:res040}.

To assess parameter estimation accuracy, we found the posterior mean squared error (PMSE) for each of the model parameters, $\alpha$, $\beta$, and $\sigma^2$. The PMSE of variable $\eta$ is the mean squared difference between the MCMC draws $\eta^{(i)}$ and the true parameter value $\eta_{\text{true}}$
\begin{align*}
    \text{PMSE}(\eta) &= \int (\eta - \eta_{\text{true}})^2\pi(\eta | \{\mathbf{r}\}_\text{obs})d\eta\\
    &\approx\frac{1}{100,000} \sum_{i=1}^{100,000} (\eta^{(i)} - \eta_{\text{true}})^2.
\end{align*}
We constructed 95\% equal-tailed credible intervals for $\alpha$, $\beta$, and $\sigma^2$ and recorded credible interval width to assess parameter estimation precision.

To assess prediction accuracy for missing time points, mean squared predictive errors (MSPE) were found for each subsample, where
\[\text{MSPE}(\{\textbf{r}\}_{\text{unobs}}) = \sum_{k\in \{\tau: \ \mathbf{r}_\tau \in \{\mathbf{r}\}_{\text{unobs}}\}} \left[ \bar{\mathbf{r}}_k^{(i)} - \mathbf{r}_k\right]'\left[ \bar{\mathbf{r}}_k^{(i)} - \mathbf{r}_k\right],\]
$\mathbf{r}_k$ is the true location of observation $k$, $\mathbf{r}_k^{(i)}$ is the $i^\text{th}$ sample from the posterior distribution of $\mathbf{r}_k$, and $\bar{\mathbf{r}}_k^{(i)} = \frac{\sum_{i=1}^{100,000}\mathbf{r}_k^{(i)}}{100,000}$. We found the mean width of 95\% equal-tailed credible intervals for missing values $\{\textbf{r}\}_{\text{unobs}}$ to assess precision of the predictions.

Figure \ref{fig:res030} depicts the statistics for all converged simulations (that is, all with Geweke $Z$-scores for $\alpha$, $\beta$, and $\sigma^2$ less then 3), and Figure \ref{fig:res040} looks at a further subset, the "best case" simulations (those which included the true values of $\alpha$, $\beta$, and $\sigma^2$ in their equal-tailed 95\% credible intervals). The results displayed in Figure \ref{fig:res030} indicate the LARI subsamples outperform the regular subsamples with respect to 95\% credible interval width and PMSE for $\alpha$, $\beta$, and $\sigma^2$. The LARI subsamples also outperform the regular subsamples on average when we compare them based on the metrics for predicting missing data. However, in the "best case" subset where LARI and regular subsamples both estimate the parameters well, the regular subsamples more accurately predict missing data. The missing data MSPEs from all but one of the LARI subsamples are lower than all MSPEs from the regular subsamples.

Thus, for the simulations that converged, the LARI sampling design led to better estimation of the model parameters $\alpha$, $\beta$, and $\sigma$ as well as better prediction of missing locations; but when both LARI and regular subsamples estimated the parameters well, the regular sampling design led to greater accuracy and precision in prediction of missing data points. This result is consistent with the hypothesis that the variability in time intervals between observations in the LARI design leads to a better understanding of movement behavior through greater accuracy in model parameter estimation. 

\FloatBarrier

\FloatBarrier

\begin{figure}[h]
\centering\includegraphics[width=\linewidth]{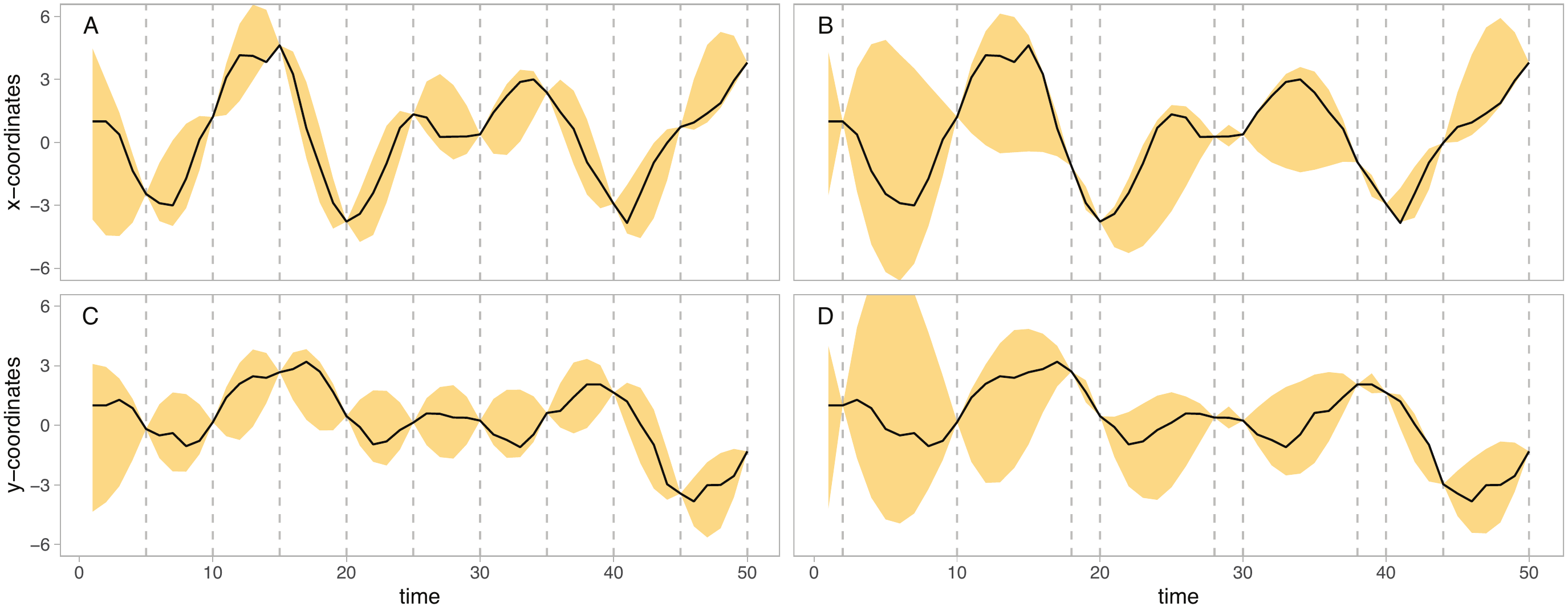}
\caption{The first 50 time points with the true simulated values connected by a black line and 95\% credible intervals depicted as orange bands. Grey dashed lines indicate the time points that were observed in the sample. The panels on the left (A and C) depict the results for the regular sample and the panels on the right (B and D) depict the results for the LARI sample.}
\label{fig:res020}
\end{figure}

To better understand the missing data prediction results, we explore one of the "best case" simulations. In Figure \ref{fig:res020}, we plot the true $x$- and $y$-coordinates for the first 50 time points with corresponding 95\% credible intervals. As shown in Figure \ref{fig:res020}, sampling at regular intervals often results in smaller credible intervals for unobserved values. We suspect this is because the LARI design includes larger time gaps than regular sampling, which disproportionately affects the mean of credible interval widths.

\FloatBarrier

\section{Guppy Data Example}\label{sec:gup}

In our first data example, we will use movement data from a captive population of guppies (\textit{Poecilia reticulata}). The group of guppies were released in the bottom right corner of a flat-bottomed square tank and swam toward a sheltered area in the opposite corner of the tank. The data consists of 360 observations recorded at 0.1 second intervals for each of 10 guppies. For more information regarding data collection, see \citet{bode2012distinguishing}. After fitting a SDE to the full data, we will compare the sampling schemes in \eqref{eq:sam010} and \eqref{eq:sam020} by subsampling the data and comparing the resulting model fits.

\subsection{SDE Model}

As in the simulation example, we can represent the movement of individual guppies with a set of SDEs. In this example, we define motility and potential surfaces
\begin{align*}
m(\textbf{r}_\tau)&\equiv 1\\
p(\mathbf{r}_\tau)&\equiv k|\mathbf{r}_\tau-\mathbf{a}|
\end{align*}
where $k\in  \mathbb{R}$ controls the strength of the drift toward a known point of attraction $\mathbf{a} = \begin{bmatrix}
    281&434
\end{bmatrix}'$ in the sheltered corner of the tank. The potential surface is defined in this way to elicit a constant force toward the point of attraction. This specification of potential and motility surfaces results in mean drift and magnitude of stochasticity
\begin{align*}
    \boldsymbol{\mu}(\mathbf{r}_\tau) &= -\triangledown p(\mathbf{r}_\tau) = -k\times \text{sign}(\mathbf{r}_\tau - \mathbf{a})\\
    c(\mathbf{r}_\tau) &= \sigma. 
\end{align*}
The set of SDEs \eqref{eq:sde010} and \eqref{eq:sde020} become
\begin{align}
    d\mathbf{r}_t&=\mathbf{v}_t dt\label{guppysde1}\\
    d\mathbf{v}_t&=-\beta[\mathbf{v}_t-(-k\times \text{sign}(\mathbf{r}_\tau - \mathbf{a}))] dt+\sigma\mathbf{I}d\mathbf{w}_t\label{guppysde2}
    \end{align}
and the numerical approximation \eqref{eq:num030} becomes
\begin{align}
    {\frac{\mathbf{r}_{\tau+2}-\mathbf{r}_{\tau+1}}{h_{\tau+1}}-\frac{\mathbf{r}_{\tau+1}-\mathbf{r}_{\tau}}{h_\tau}}&=\beta h_\tau \left(-k\times \text{sign}(\mathbf{r}_\tau - \mathbf{a})-\frac{\textbf{r}_{\tau+1}-\mathbf{r}_\tau}{h_\tau}\right)+h_\tau^{\sfrac{1}{2}}\sigma \boldsymbol{\epsilon}_\tau.\label{eq:guppy050}
\end{align}
Linear regression was implemented to estimate $k$, $\beta$, and $\sigma^2$.

\subsection{Results}

We implemented a data subsampling procedure which resulted in 3 regular subsamples and 300 LARI subsamples. The regular subsamples were recorded at 0.3 second intervals beginning at each of the first 3 time points. For each of the regular subsamples, 100 corresponding LARI subsamples were collected which consisted of every other sample from the regular data (i.e., regular samples at 0.6 second intervals) coupled with a random time point selected from the observations in each subinterval. This resulted in 3 groups of subsamples, each consisting of one regular subsample and 100 corresponding LARI subsamples.

\begin{figure}
\centering\includegraphics[width=.9\linewidth]{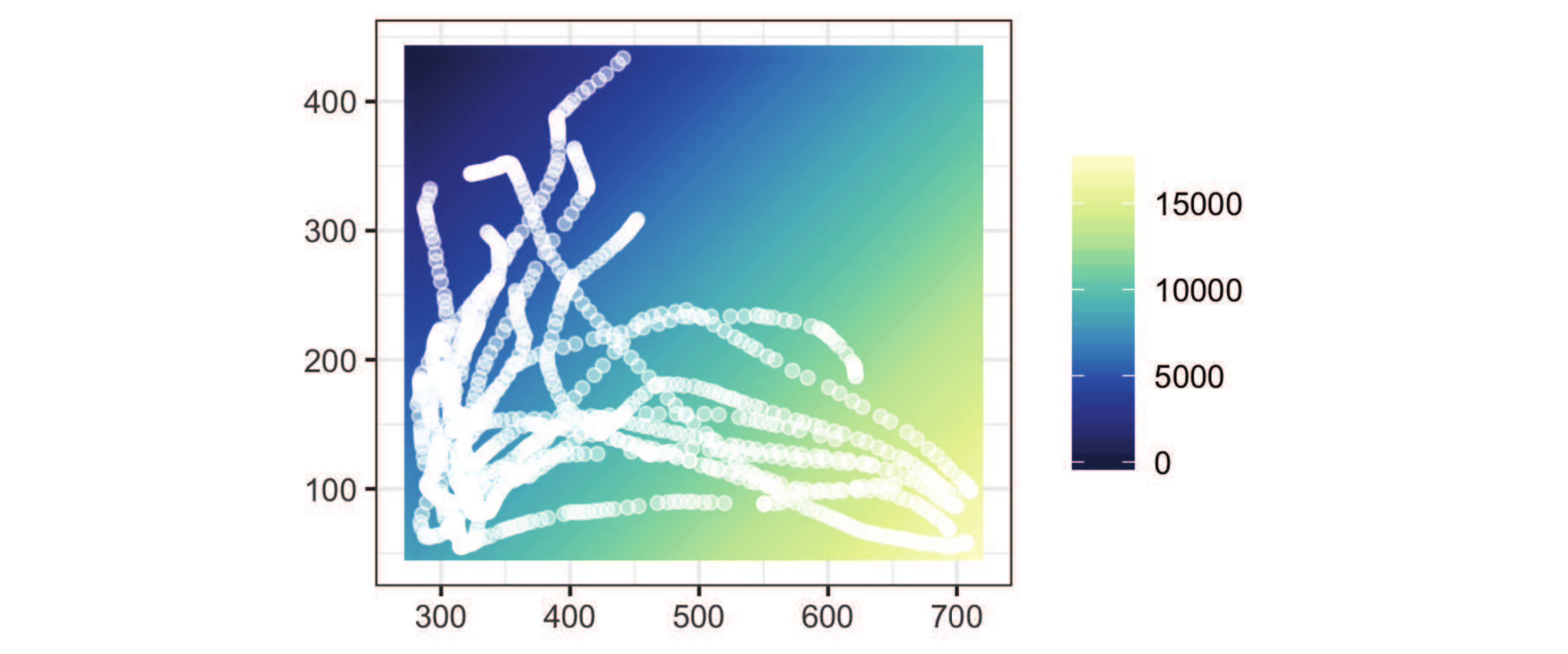}
\caption{The estimated potential surface using the full data. Observations are shown in white.}
\label{fig:gup010}
\end{figure}

\begin{figure}
\centering\includegraphics[width=\linewidth]{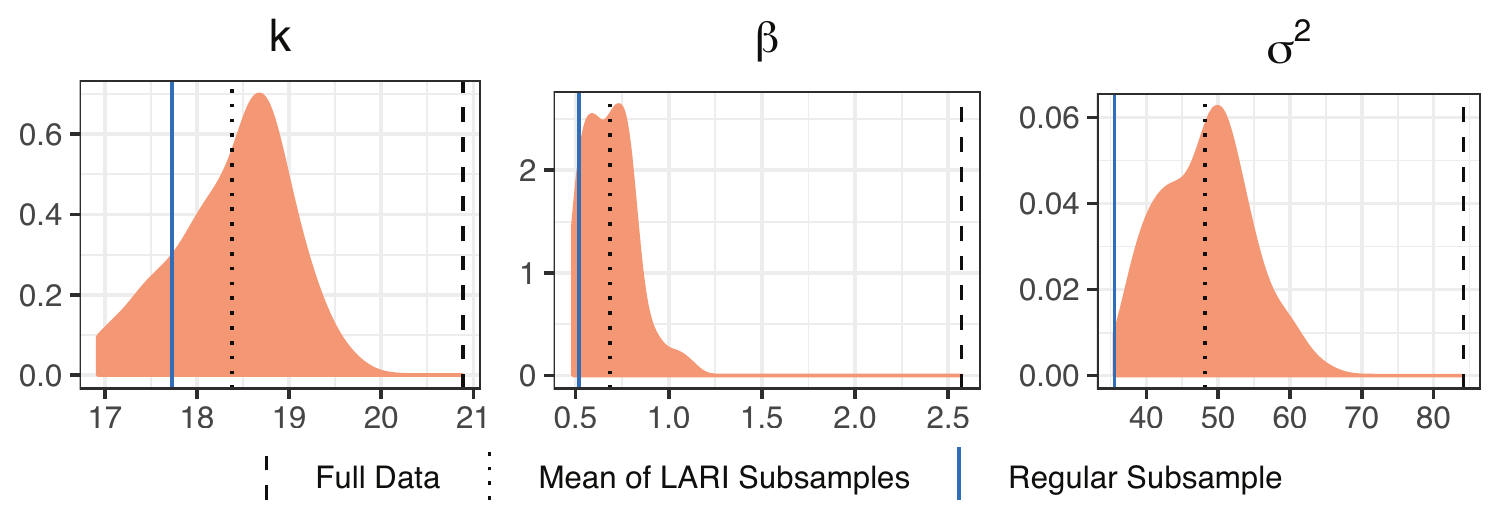}
\caption{The estimated values of the parameters $k$, $\beta$ and $\sigma^2$ from a regular subsample and 100 corresponding LARI subsamples (orange empirical density).}
\label{fig:gup020}
\end{figure}

The potential surface estimated with the full data is shown in Figure \ref{fig:gup010} with observations depicted in white. The potential surfaces estimated with subsampled data are similar in appearance, so we analyze them by comparing estimates of $k$, $\beta$, and $\sigma^2$. The 3 groups of subsamples resulted in identical conclusions, so we chose a random group to plot in Figure \ref{fig:gup020}. In Figure \ref{fig:gup020}, we display the parameter estimation results for $k$, $\beta$, and $\sigma^2$. While all subsamples led to underestimation of the model parameters compared to the full data, the LARI subsamples are closer to the full data. In particular, when estimating $\sigma^2$, the LARI subsamples always outperform the regular subsample.

\FloatBarrier

\FloatBarrier

\section{Carpenter Ant Example}\label{sec:com}

\FloatBarrier

\begin{figure}[h]
\centering\includegraphics[width=.9\linewidth]{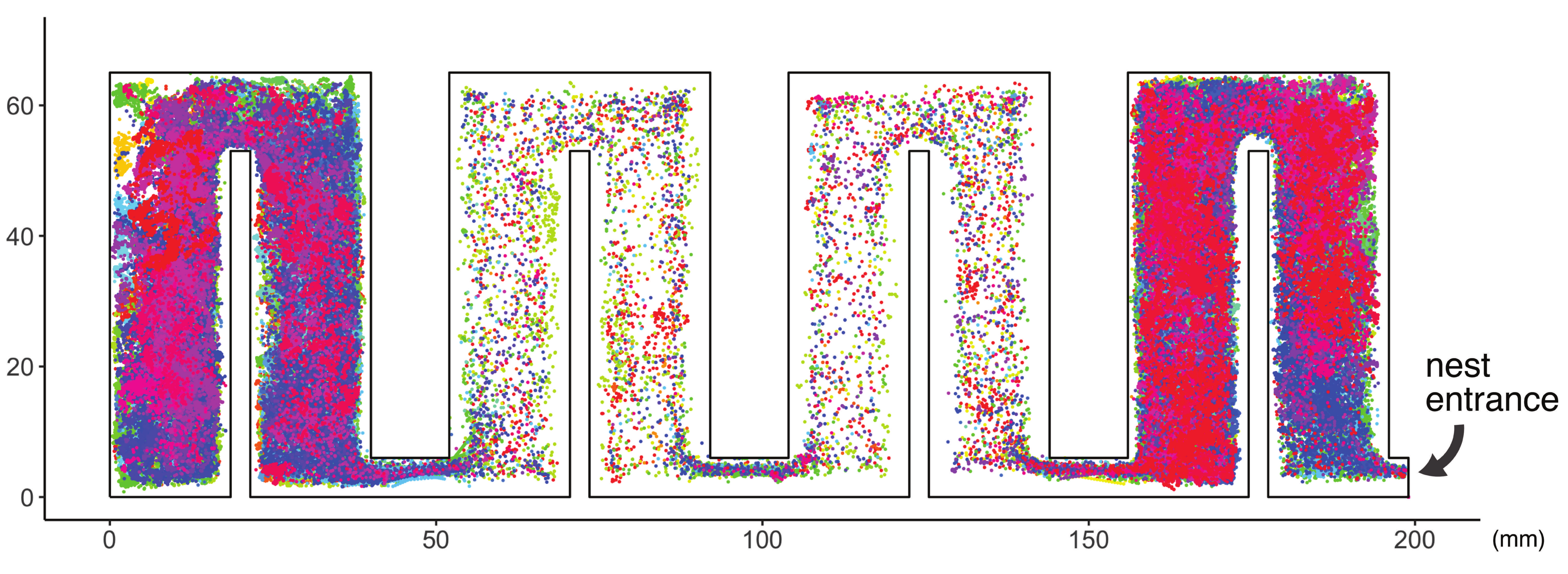}
\caption{$14,401$ positions over 4 hours for each of the $73$ ants, color coded by individual.}
\label{fig:com010}
\end{figure}

\FloatBarrier

\begin{figure}[h]
\centering\includegraphics[width=\linewidth]{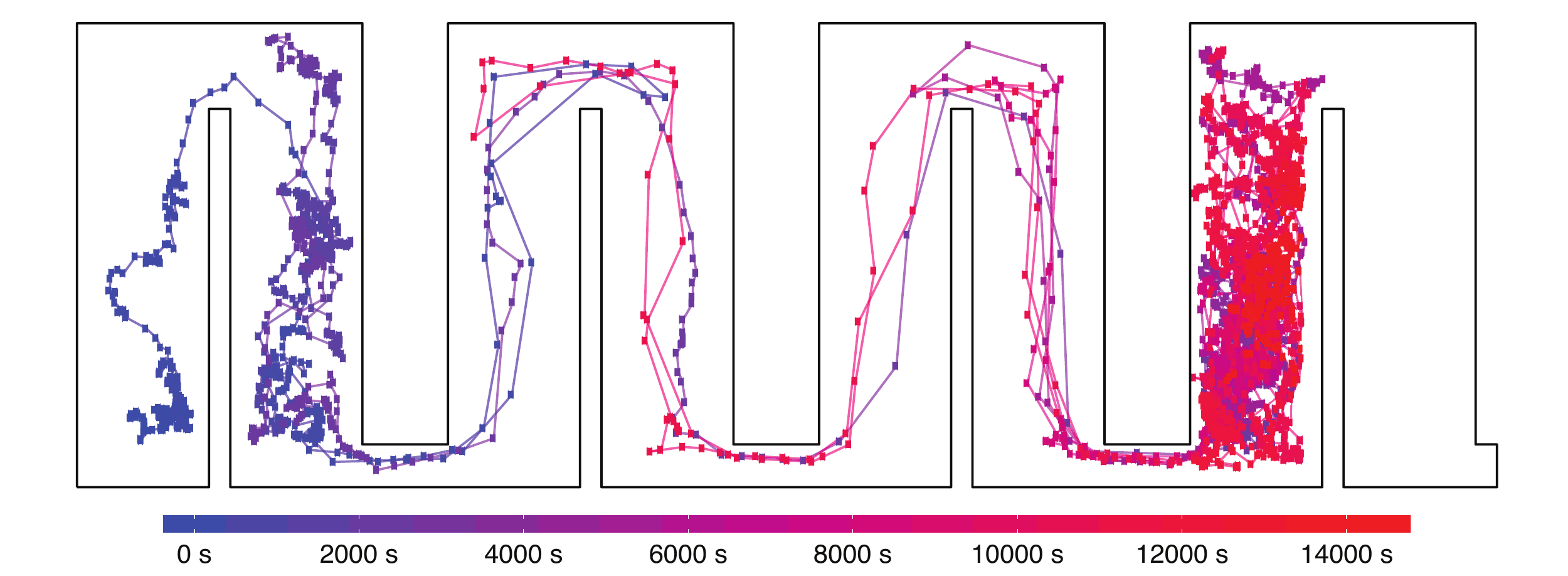}
\caption{$14,401$ positions over 4 hours for a single ant, color coded by observation time.}
\label{fig:com020}
\end{figure}

We now turn to the dataset introduced in Section \ref{sec:int}, which consists of the positions of 78 ants at 1 second intervals over a 4 hour time frame (14,401 total observations per ant). Researchers observed the ants in a $200 \times 65 \times 6$ mm wooden nest, shown in Figure \ref{fig:com010} along with the positions of all ants at all time points. The ants could enter or exit the nest at any time to utilize a separate foraging area. The data collection procedure is described in further detail by \citet{modlmeier2019ant}. In Figure \ref{fig:com020}, we illustrate movement observed for one individual who stayed inside the nest throughout the 4 hour time frame.

From this data we obtained four datasets for comparison, one of which is the full data (1 second intervals). The three additional datasets are subsamples from the full data; we produced one with the regular sampling design using $h = 3$ seconds, one dataset with the LARI sampling design using $h=5$ seconds, and one with the regular sampling design using $h = 5$ seconds. Ants display stop and start behavior, but the SDE model alone cannot handle state switching. Thus, we removed observations where the ants were stationary within each dataset. After removal of stationary observations, a total of $232,571$ observations remained in the full dataset, the largest of the four datasets. Similar to \citet{russell2018}, we only consider modeling ant movement when ants are moving.

\FloatBarrier

\subsection{Ant Movement Model}\label{sec:antmodel}

In this example, we apply the framework from \eqref{eq:num030} to model ant movement behavior. We represent the surface of the ant nest using $J=9,998$ grid cells ($1\times 1$ mm). Following \citet{russell2018}, we specify spatially-varying motility and potential surfaces to capture spatial heterogeneity in ant movement. The zeroth order spline representations of the potential and motility surfaces evaluated at position $\mathbf{r}_\tau$ are
\begin{align*}
    p(\mathbf{r}_\tau)&\equiv \sum_{j=1}^J p_j s_j(\mathbf{r}_\tau)\\
    m(\mathbf{r}_\tau)&\equiv \sum_{j=1}^J m_j s_j(\mathbf{r}_\tau)
\end{align*}
where \[s_j(\mathbf{r}_\tau)\equiv
\begin{cases}
1, & \mathbf{r}_\tau \text{ in }j^{th}\text{ grid cell}\\
0, & \text{otherwise}
\end{cases}\] and $p_j$ and $m_j$ are the potential and motility surfaces respectively, evaluated in grid cell $j$. Of course, there are other basis functions we could use to build potential and motility surfaces, such as thin plate splines. However, thin plate splines and other bases are more difficult work with in the constrained geometry of the ant nest.

The model equation \eqref{eq:num030} has infinitely many solutions if constraints are not imposed. To obtain identifiability, we fix $\sigma=1$ as in \citet{russell2018}. \citet{russell2018} took a Bayesian approach to parameter estimation with a similar model. However, our novel approximation in \eqref{eq:10} allows for a direct evaluation of the likelihood of animal locations observed at irregular time intervals. We propose an algorithm for estimation of model parameters based on maximizing the likelihood \eqref{eq:10} while penalizing the roughness of the potential and motility surfaces. \citet{modlmeier2019ant} used a related algorithm, but only allowing for regularly sampled data. \citet{modlmeier2019ant} also do not provide full mathematical details of the algorithm, which we provide in summary here and in detail in Supplemental Materials \ref{sec:antest}.

We estimate $\mathbf{p} \equiv \begin{bmatrix}
    p_1 & \hdots & p_J
\end{bmatrix}'$ and $\mathbf{m} \equiv \begin{bmatrix}
    m_1 & \hdots & m_J
\end{bmatrix}'$ with an iterative procedure beginning with the model equation \eqref{eq:num030}. Rewriting \eqref{eq:num030}, we have
\begin{align}
{\frac{\mathbf{r}_{\tau+2}-\mathbf{r}_{\tau+1}}{h_{\tau+1}}-\frac{\mathbf{r}_{\tau+1}-\mathbf{r}_{\tau}}{h_\tau}}&\sim\text{N}\left(\beta h_\tau \left(\boldsymbol{\mu}(\mathbf{r}_\tau)-\frac{\mathbf{r}_{\tau+1}-\mathbf{r}_\tau}{h_\tau}\right),\text{diag}(c^2(\mathbf{r}_\tau) h_\tau) \right)
\end{align}
where $\boldsymbol{\mu}(\mathbf{r}_\tau) = m(\mathbf{r}_\tau)[-\triangledown p(\mathbf{r}_\tau)]$ is estimated as a function of $\mathbf{m}$ and $\mathbf{p}$ and $c(\mathbf{r}_\tau) = m(\mathbf{r}_\tau)$ is estimated as a function of $\mathbf{m}$. Thus, we refer to $\mathbf{p}$ as a mean parameter, and $\mathbf{m}$ could be considered both a mean and a variance parameter. However, we will estimate $\mathbf{m}$ using the variance and thus we refer to it as a variance parameter.

 We hold out 20\% of the data to use when choosing the tuning parameter $\lambda$, which controls the smoothness of the surfaces, later in the procedure. 
The remaining 80\% of the data are fit simultaneously for all ants, assuming ants move independently and there is no correlation in the x- and y-components of movement. The procedure is similar to restricted maximum likelihood (REML) approaches common in mixed models, as we use residuals to estimate covariance parameters, which are then used to estimate mean parameters. Our proposed approach is as follows:
\begin{enumerate}
    \item Obtain a preliminary estimate of mean parameters ($\beta$ and $\mathbf{p}$) assuming the motility surface is constant (model errors are independent and identically distributed).
    \item Estimate variance parameters ($\mathbf{m}$) using residuals from step 1.
    \item Estimate mean parameters ($\beta$ and $\mathbf{p}$) conditioned on the variance estimates from step 2.
\end{enumerate}
For details about the estimation approach, refer to Supplemental Material \ref{sec:antest}.

\subsection{Results}

Section \ref{sec:antmodel} describes a computationally efficient method of fitting spatially-varying coefficients in SDE movement models. We completed the 3-step procedure for 17 values of the tuning parameter using the full data (232,571 total observations) in less than 25 minutes. We completed the procedure in the programming language R (version 3.5.2) on a MacBook Pro with a 2.9 GHz Intel Core i5 processor and 8 GB of 2133 MHz LPDDR3 RAM.

We estimated motility and potential surfaces for four samples: the full data with observations at 1 second intervals, a subsample with observations every 3 seconds, a subsample with observations every 5 seconds, and a LARI subsample with regular samples every 10 seconds coupled with a random point in between each pair of regular samples. The 10 second LARI sample and every 5 second regular sample have an equal number of observations, so comparison of the results from these two datasets amounts to a direct comparison of the LARI sampling scheme to the regular sampling scheme for this data.

For each of the four samples, we chose the optimal value of $\lambda$ separately based on prediction accuracy on the holdout set. We chose log$(\lambda) = 0$ for the full data, log$(\lambda) = 2$ for the every 3 seconds sample, log$(\lambda) = 3$ for the every 5 seconds sample, and log$(\lambda) = 4$ for the 10 second LARI sample.

\begin{figure}[h]
\centering\includegraphics[width=\linewidth]{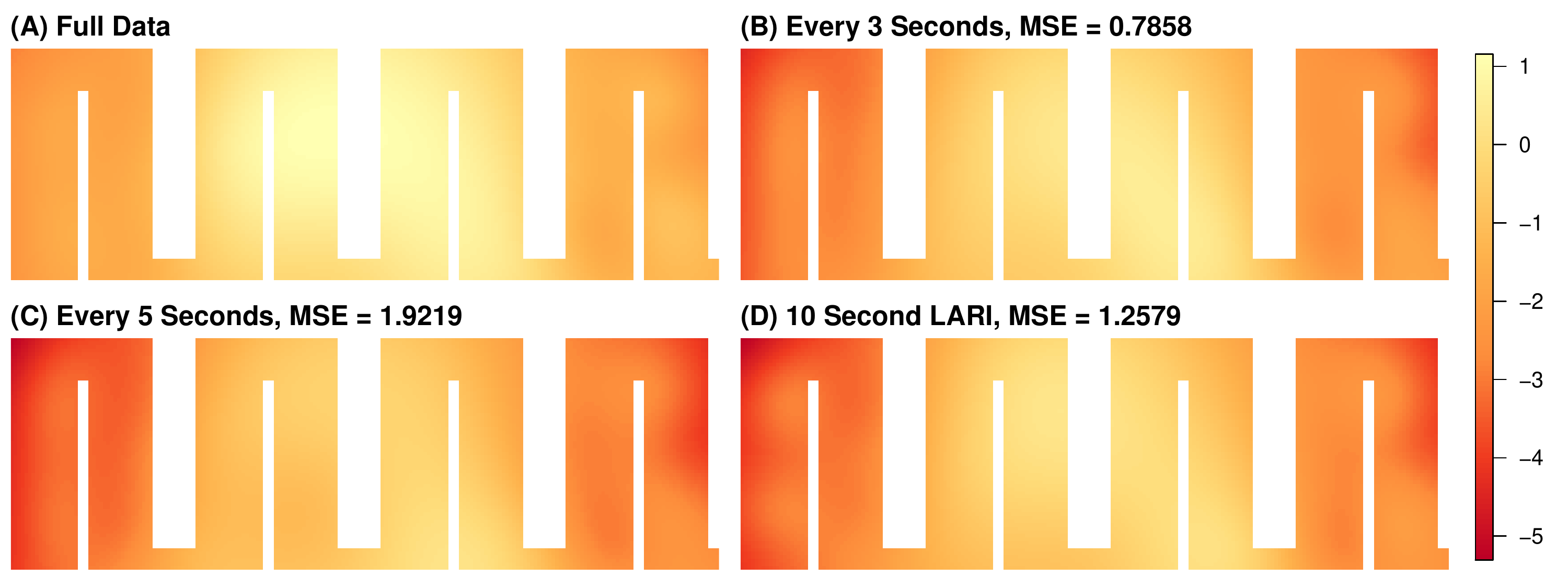}
\caption{The natural log of the motility surfaces estimated using the full data (A) and 3 subsamples (B)-(D).}
\label{fig:comres010}
\end{figure}

\begin{figure}[h]
\centering\includegraphics[width=\linewidth]{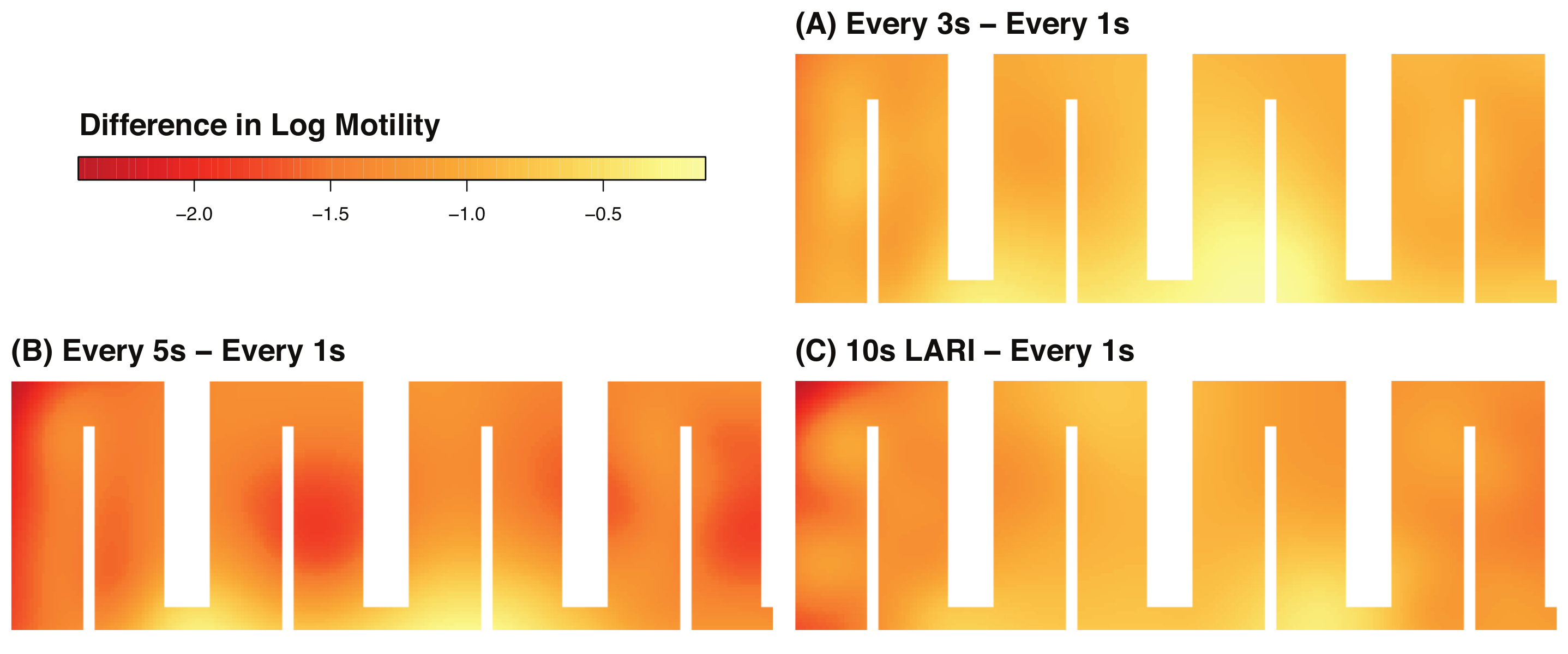}
\caption{Differences between the estimated log motility surfaces from each of the subsamples and the full data (calculated by grid cell). Negative values indicate underestimation of the motility surface.}
\label{fig:comres020}
\end{figure}

Figure \ref{fig:comres010} displays the estimated log motility surfaces for the four datasets. Since high motility indicates high activity, it is evident from the plots that the ants moved more quickly in the center chambers \citep{modlmeier2019ant}. Assuming the motility surface generated with the full data is closest to the truth, we found mean squared errors (MSE) of the log motility surfaces in Figure \ref{fig:comres010}(B--D) by summing squared differences between those surfaces and the surface in Figure \ref{fig:comres010}(A) over all grid cells. As shown in the text in Figure \ref{fig:comres010}, the MSE for the 10 second LARI sampling scheme is smaller than the every 5 seconds sampling scheme, which suggests that the motility surface was better estimated with the LARI subsample than the regular subsample of the same size. Each of the three subsamples underestimated the motility surface compared to the full data, a sign that regardless of sampling design, we lost information about the motility surface by subsampling (Figure \ref{fig:comres020}).

\begin{figure}[h]
\centering\includegraphics[width=\linewidth]{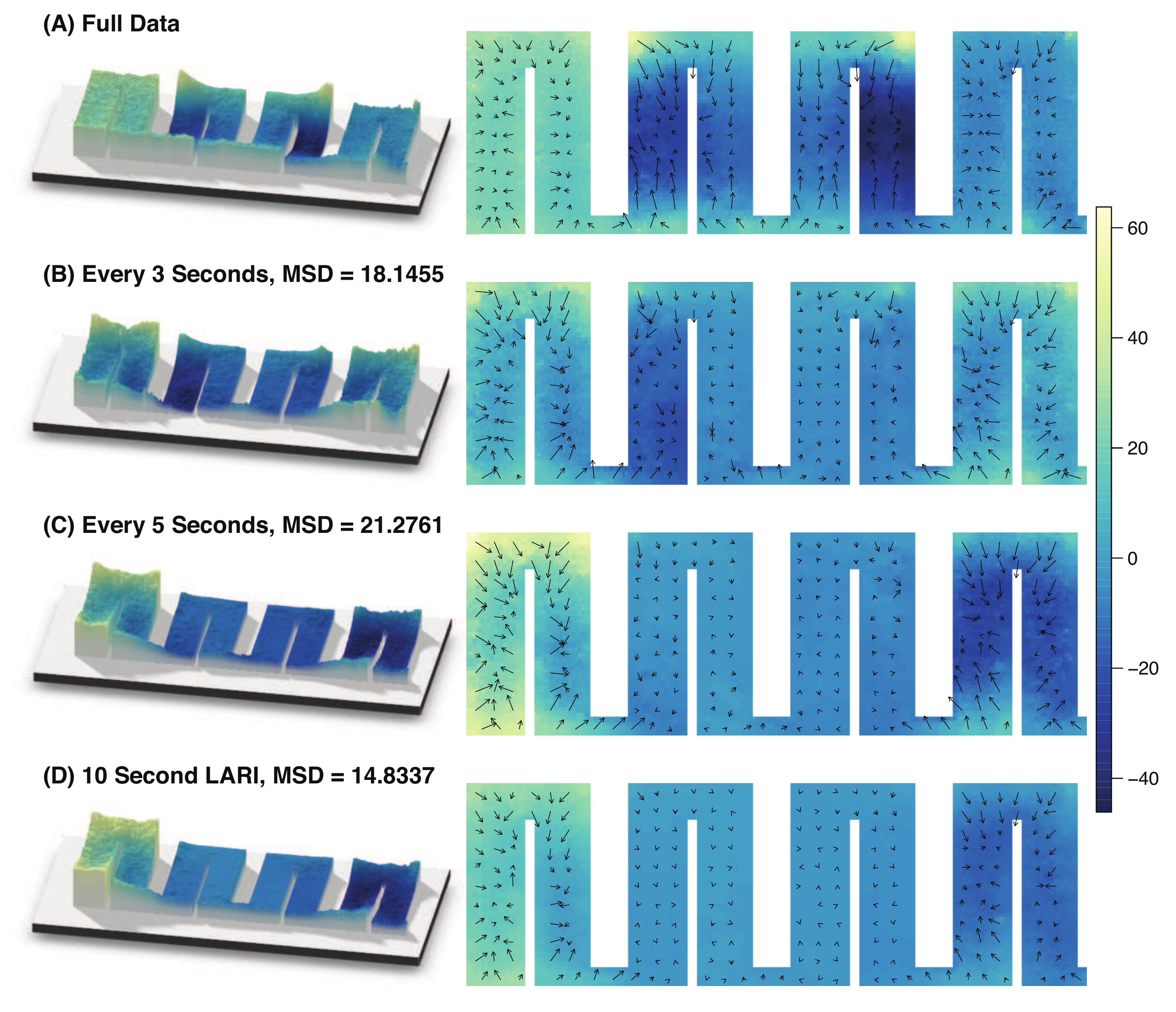}
\caption{Potential surfaces estimated with the four samples. The same potential surfaces are plotted in three dimensions on the left (using the \pkg{rayshader} R package) and in two dimensions on the right.}
\label{fig:comres030}
\end{figure}

In Figure \ref{fig:comres030}, we show the estimated potential surfaces with gradient vectors pointing down the gradient of potential surface. The gradient vectors depict the negative gradient of the potential surface scaled by 5 to improve visibility. We chose to plot the gradient vectors in every third grid cell for visual clarity. Since the potential surface is identifiable only up to an additive constant, we subtracted the mean from each potential surface to view them on roughly the same scale. We carried out comparisons between potential surfaces through mean squared distance (MSD) between the ends of the gradient vectors generated from the full data and those generated from the subsample. Of the three subsamples, the potential surface estimated with the 10 second LARI subsample had the smallest MSD (Table \ref{tab:poterrors}). We then decomposed the MSD into two additional metrics: mean error in magnitude and angle of the gradient vectors. The 10 second LARI design resulted in a smaller error in the angle of the gradient vectors compared to the every 5 second design, but the 5 second design resulted in a smaller mean error in gradient vector magnitude (Table \ref{tab:poterrors}). On average, all three potential surfaces estimated with subsamples of the data are too smooth, i.e., the gradient vectors are biased toward zero (Table \ref{tab:poterrors}).

\begin{table}[]
\centering
\begin{tabular}{llll}
\hline
 & Every 3 Seconds & Every 5 Seconds & 10s LARI \\ \hline
Mean Error in Gradient Vector Magnitude & -0.2216 & -0.5045 & -1.624 \\

Mean Error in Gradient Vector Angle & -0.0317 & 0.0565 & 0.0164 \\
MSD between Gradient Vectors & 18.1455 & 21.2761 & 14.8337
\end{tabular}
\caption{Potential surface error statistics for the three subsamples.}
\label{tab:poterrors}
\end{table}

\begin{figure}[h]
\centering\includegraphics[width=\linewidth]{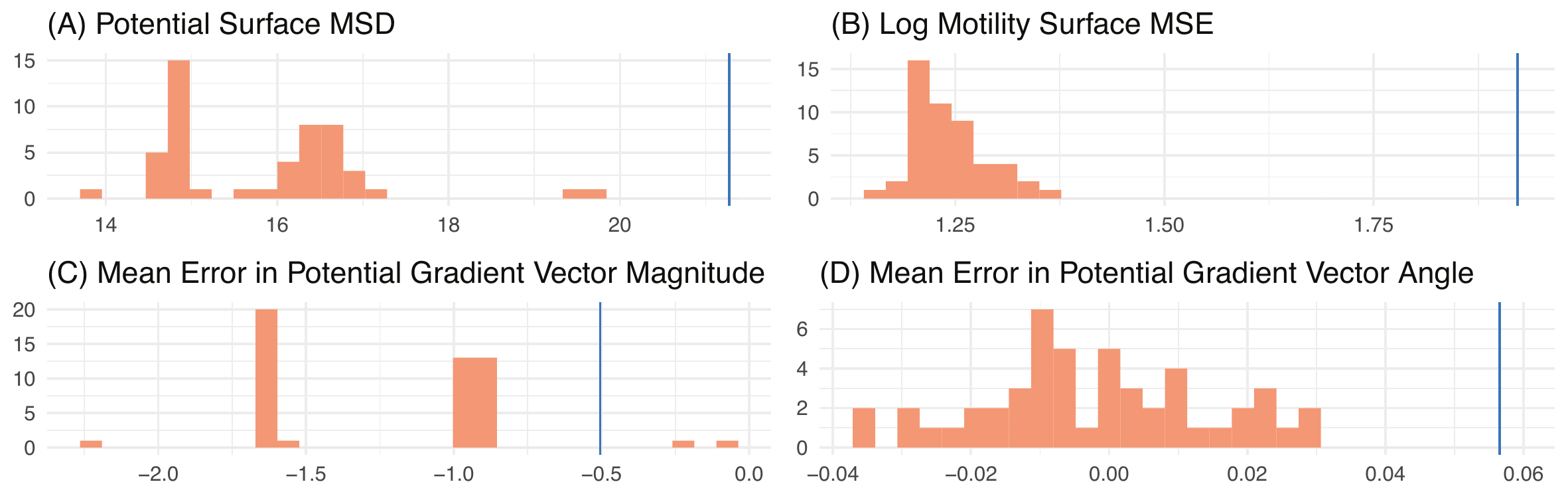}
\caption{Statistics calculated for the motility and potential surfaces fit using 50 different 10 second LARI subsamples (orange) are compared to statistics from the every 5 second subsample (blue).}
\label{fig:comres040}
\end{figure}

Since the LARI sampling scheme requires random samples in each 10 second time interval, there are $9^{1,439}$ possible 10 second LARI subsamples. To evaluate the variability attributed to this random component, we took 50 different 10 second LARI subsamples and fit each subsample separately. We found that the 10 second LARI subsample consistently outperformed the every 5 second subsample in regard to all the metrics we looked at except magnitude of potential surface gradient vectors (Figure \ref{fig:comres040}).

\FloatBarrier

\FloatBarrier

\FloatBarrier

\section{Discussion}

The simulation, guppy data, and ant data examples describe vastly different systems, but the sampling procedures laid out in Section \ref{sec:sam} and general model framework described in Section \ref{sec:sto} were applicable in all three cases. We chose to highlight these three examples to emphasize the generalizability of the SDE framework and our proposed sampling approach. In all of the examples, the LARI sampling design led to greater accuracy in parameter estimation compared to samples at regular time intervals. As shown in the simulation example in Section \ref{sec:sim}, the LARI subsample also resulted in better prediction of missing values compared to the regular subsample. We conclude that when conducting animal movement research on data similar to that examined in this paper, a LARI sample is preferable to a regular sample of the same sample size and duration.


We determined that predicting finer scale movement (infill) was greatly useful for parameter estimation in the simulation example (see Supporting Materials \ref{sec:appendix:sim} for details). This result implies that imputation of missing data at a finer scale might improve parameter estimation in the guppy and ant data examples as well. In the ant example, we found that the motility surfaces were underestimated using both regular and LARI sampling schemes. Augmenting observations with additional latent infill points in the ant example would introduce more variation in the movement paths, potentially reducing the underestimation of the motility surface. In the guppy example, all three parameters were underestimated when the data were subsampled, suggestion that augmenting observations might be useful here as well. However, in the simulation example, predicting missing data with the Metropolis-within-Gibbs algorithm was computationally intensive. We needed 90 hours of computational time on a high-performance computing cluster to simulate, subsample, and fit each of the 150 datasets.

While computational complexity was an issue in the simulation example, the multi-step model-fitting procedure in the ant example was extremely computationally efficient. For each sample, less than 25 minutes on a laptop computer were required to fit the model with a range of 17 different tuning parameter values. The scalability and computational feasibility of adding components to this model are huge assets.

In this paper, we presented a general SDE modeling framework along with three model-fitting procedures. The SDE framework has a wide range of possible extensions, including the addition of known seasonal variation, asynchronous movement across individuals, and state switching.
The model framework as presented here assumes the animals are in constant motion throughout the study period, allowing us to describe movement behavior while an individual is in motion but not when the individual is stationary. In the ant example, we met this restriction by removing all data points where the ants were not moving. The addition of state switching would allow us to capture the start and stop behavior in the ant data and could be used to predict finer scale movement.

In this paper, we have compared two sampling designs in the context of animal movement. Thus, we have barely scraped the surface of a research area which is largely unexplored: optimal sampling for animal movement. As of present, there is no comprehensive guide to sampling design for animal movement. A thorough examination of sampling design for animal movement would allow researchers to allocate resources more efficiently and discover details of animal movement behavior that might otherwise be lost.

\newpage

\section*{Funding information:}

NSF EEID 1414296, NIH GM116927-01

\section*{Data availability}

The raw carpenter ant data is available through Dryad (DOI: 10.5061/dryad.sh4m4s6), as is the guppy data (DOI: 10.5061/dryad.kt3109v7).

\bibliography{bib1}
\bibliographystyle{apalike}

\newpage
\pagenumbering{arabic}
    \setcounter{page}{1}

\title{Supplement to A Lattice and Random Intermediate Point Sampling Design for Animal Movement}
\maketitle

\appendix

\section{Examples of Model Capabilities}\label{sec:appendix:ex}

\begin{figure}[h]
\centering\includegraphics[width=\linewidth]{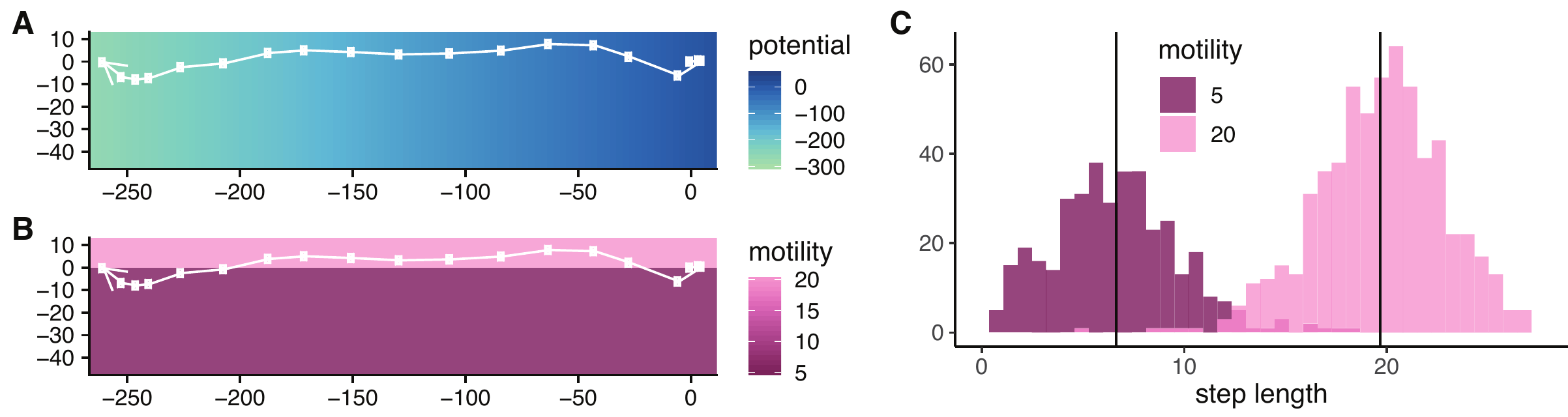}
\caption{Panels (A)-(B): 20 locations simulated from \eqref{eq:10} with parameters \eqref{eq:op1a}--\eqref{eq:op1b}. The simulations are shown with the potential surface (A) and motility surface (B). The arrow points in the direction the simulated individual is heading. Panel (C): 1,000 locations simulated from \eqref{eq:10} with parameters \eqref{eq:op1a}--\eqref{eq:op1b}. Panel (C) displays histograms of the distances from $\mathbf{r}_{\tau - 1}$ to $\mathbf{r}_\tau$ for each possible value of $m(\mathbf{r}_{\tau - 2})$. Black vertical lines represent the means for each group.}
\label{fig:op1}
\end{figure}

We will briefly examine data simulated from \eqref{eq:10} with 3 different sets of parameters. In all 3 simulations, we started the simulation at $\begin{bmatrix}     0&0 \end{bmatrix}'$ for the first 2 time steps and used $\beta = 0.4$ and $\sigma = 0.5$. In the first simulation,
\begin{align}
    p(\mathbf{r}_\tau) &= x_\tau\label{eq:op1a}\\
    m(\mathbf{r}_\tau) &= \begin{cases}
        5, & y_\tau \leq 0\\
        20, & y_\tau > 0
    \end{cases}\label{eq:op1b}
\end{align}
result in movement with average drift in the negative $x$ direction and where the average speed is higher in the space where $y > 0$ than where $y \leq 0$. Since in \eqref{eq:10} each simulated value $\mathbf{r}_\tau$ is a function of the motility evaluated two observations prior, the average speed at $\mathbf{r}_\tau$ is higher (and step size smaller) when $y_{\tau - 2} > 0$ than when $y_{\tau - 2} \leq 0$. The first 20 simulated time points are plotted in Figure \ref{fig:op1}A-B with the potential and motility surfaces which generated them. Figure \ref{fig:op1}C depicts the difference in average step size attributed to the motility surface in a simulation of 1,000 time steps.

\FloatBarrier
\FloatBarrier

\begin{figure}[h]
\centering\includegraphics[width=\linewidth]{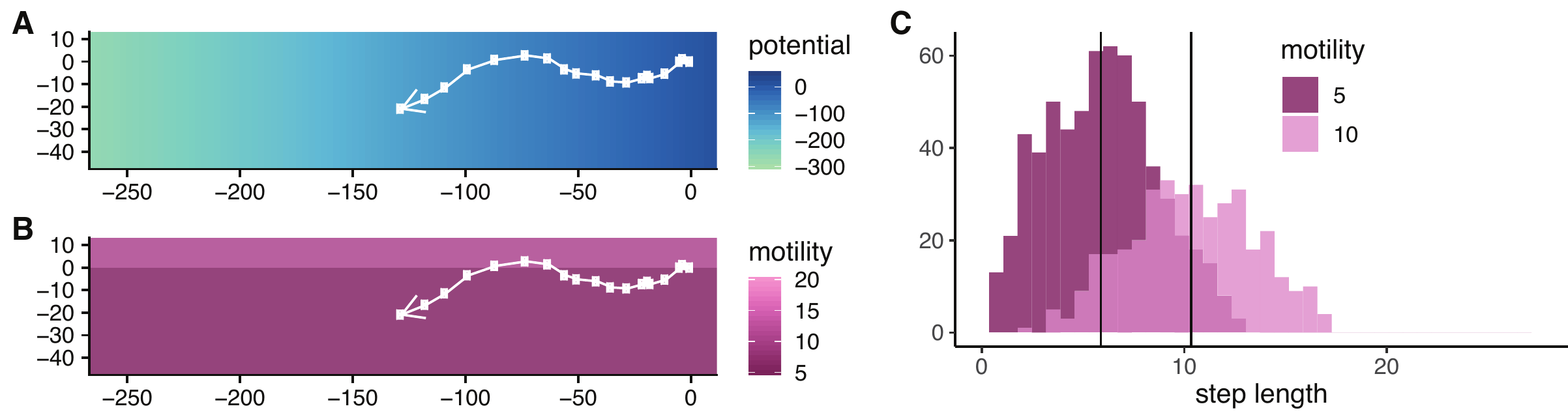}
\caption{Plots are as in Figure \ref{fig:op1} using parameters \eqref{eq:op1a} and \eqref{eq:op3}.}
\label{fig:op3}
\end{figure}

We reran the simulation with
\begin{align}
    m(\mathbf{r}_\tau) &= \begin{cases}
        5, & y_\tau \leq 0,\\
        10, & y_\tau > 0.
    \end{cases}\label{eq:op3}
\end{align}
This change left the average step size in the low motility area unchanged but led to a reduction in the average step size in the high motility area (Figure \ref{fig:op3}).

\FloatBarrier
\FloatBarrier

\begin{figure}[h]
\centering\includegraphics[width=\linewidth]{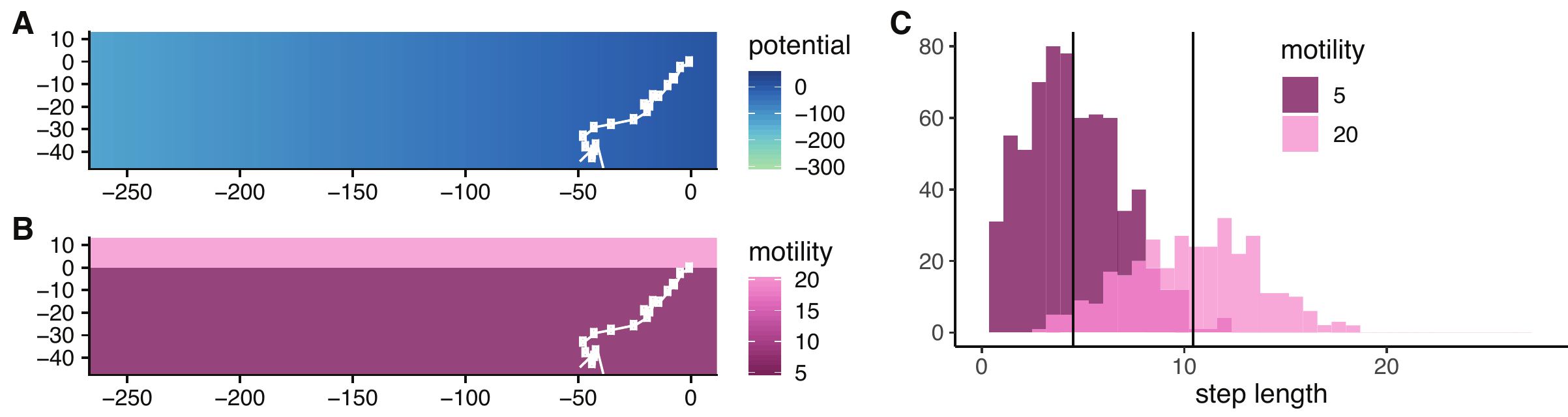}
\caption{Plots are as in Figure \ref{fig:op1} using parameters \eqref{eq:op2} and \eqref{eq:op1b}.}
\label{fig:op2}
\end{figure}

Lastly, we used the original motility surface \eqref{eq:op1b} and reduced the severity of the gradient of the potential surface by letting
\begin{align}
    p(\mathbf{r}_\tau) &= 0.5x_\tau \label{eq:op2}
\end{align}
which led to a reduced average step size in both groups (Figure \ref{fig:op2}C) as well as less direct movement down the gradient of the potential surface (Figure \ref{fig:op2}A-B).

\FloatBarrier
\FloatBarrier

\section{Simulation Example without Infill}\label{sec:appendix:sim}

Dividing \eqref{eq:sim050} by $h_{\tau}^{\sfrac{1}{2}}$ results in
\begin{align}
    {\frac{\mathbf{r}_{\tau+2}-\mathbf{r}_{\tau+1}}{h_{\tau+1}h_{\tau}^{\sfrac{1}{2}}}-\frac{\mathbf{r}_{\tau+1}-\mathbf{r}_{\tau}}{h_{\tau}^{\sfrac{3}{2}}}}&=\alpha \left(-2 h_\tau^{\sfrac{1}{2}} \mathbf{r}_\tau \right) - \beta \left(\frac{\textbf{r}_{\tau+1}-\mathbf{r}_\tau}{h_\tau^{\sfrac{1}{2}}}\right)+\sigma \boldsymbol{\epsilon}_\tau,\label{eq:sim070}
\end{align}
which is a linear model with independent and identically distributed errors. We could then fit \eqref{eq:sim070} with simple linear regression. Assessment of linear regression estimation accuracy was carried out on a subsample including every other simulated data value. Thus, the subsample is very similar to the true data. However, of the 95\% confidence intervals constructed for $\beta$, $\alpha$, and $\sigma$, only the confidence interval for $\beta$ captured the true parameter value. Adoption of a Bayesian approach would allow us to place priors on the parameters that may lead to better approximations.

\FloatBarrier
\begin{figure}[h]
\centering\includegraphics[width=.85\linewidth]{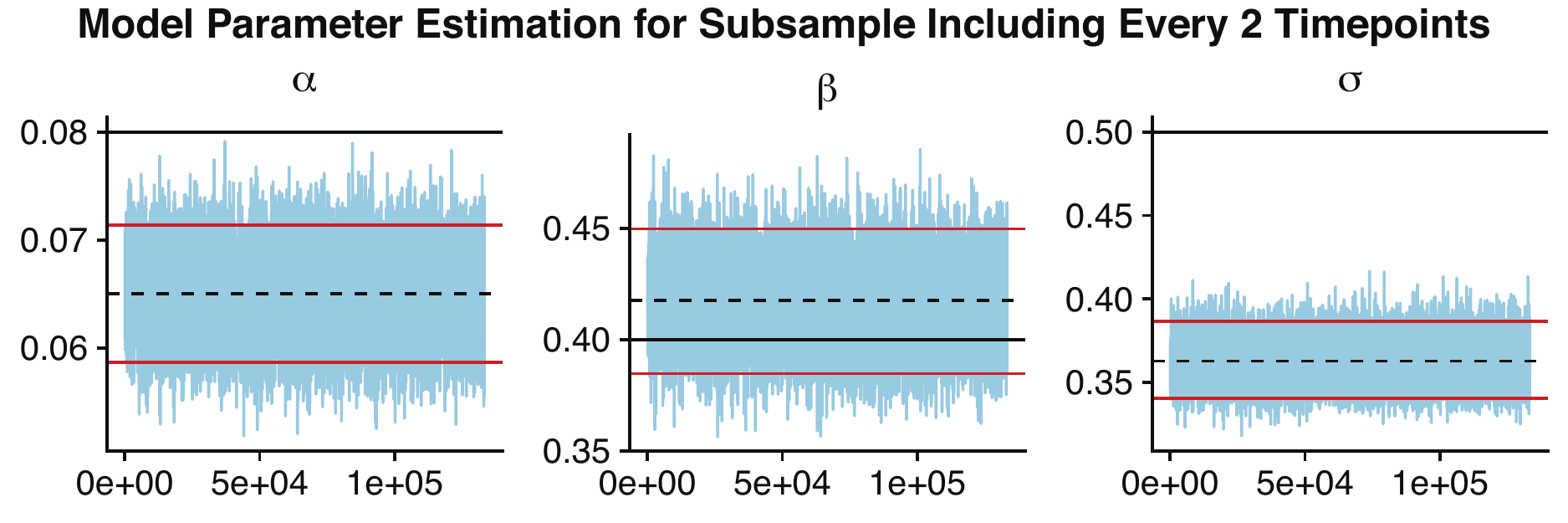}
\caption{MCMC draws from the posterior distribution conditioned on a regular subsample with every other timepoint. The MCMC draws are plotted in blue, the red lines bound equal-tailed 95\% credible intervals, the black dashed line is the estimated posterior mean and the black solid line is the true parameter value.}
\label{fig:sim020}
\end{figure}
\FloatBarrier

Vague priors \begin{align}
    \pi(\sigma)&= \text{InverseGamma}\left(1,1\right)\label{eq:sim080}\\
    \pi(\beta)&= \text{Exponential}\left(1\right)\label{eq:sim090}\\
    \pi(\alpha)&= \text{Normal}\left(0,10^2\right)\label{eq:sim100}
\end{align} were placed on the parameters and Metropolis-Hastings samples were drawn from the posterior distribution. As in the linear regression model, we subset the data to include every other time point. Despite this subset being very close to the original simulation, only $\beta$ was captured in its equal-tailed 95\% credible interval, as shown in Figure \ref{fig:sim020}.

\section{Bayesian framework}\label{sec:appendix:bay}

To estimate $\beta$, $\alpha$, $\sigma$, and $\{\mathbf{r}\}_{\text{unobs}}$, we took a Bayesian approach and constructed an MCMC algorithm to sample from the joint posterior $\pi(\alpha,\beta,\sigma,\{\mathbf r\}_{\text{unobs}}|\{\mathbf r\}_\text{obs})$ where
\begin{align*}
    \pi(\alpha,\beta,\sigma,\{\mathbf r\}_{\text{unobs}}|\{\mathbf r\}_\text{obs})&\propto
    P(\{\mathbf r\}|\beta, 
    \sigma, \alpha) \pi(\beta)\pi(\sigma)\pi(\alpha).
\end{align*}
Vague priors
\eqref{eq:sim080}--\eqref{eq:sim100}
were placed on the model parameters. We also tried increasing the variance of these prior distributions by a factor of 10, but we saw no meaningful difference in the resulting empirical posterior distributions.

The initial positions $\mathbf{r}_\tau=\begin{bmatrix}
x_\tau & y_\tau
\end{bmatrix}'$, $\tau = 1,2$ were assigned independent uniform priors 
\begin{align*}
x_1&\sim \text{Uniform}\left(\text{min}\{\mathbf x\}_\text{obs},\text{max}\{\mathbf x\}_\text{obs}\right)\\
y_1&\sim \text{Uniform}\left(\text{min}\{\mathbf y\}_\text{obs},\text{max}\{\mathbf y\}_\text{obs}\right)\\
x_2&\sim \text{Uniform}\left(\text{min}\{\mathbf x\}_\text{obs},\text{max}\{\mathbf x\}_\text{obs}\right)\\
y_2&\sim \text{Uniform}\left(\text{min}\{\mathbf y\}_\text{obs},\text{max}\{\mathbf y\}_\text{obs}\right)
\end{align*}
resulting in the joint distribution of the observed and unobserved positions
\begin{align*}
P(\{\mathbf{r}\}|\beta, \sigma, \alpha) = P(x_1)P(y_1)P(x_2)P(y_2)\prod_{\tau = 1}^{n-2} P(\mathbf r_{\tau+2}|\mathbf r_{\tau+1},\mathbf r_{\tau}, \beta, \sigma, \alpha).
\end{align*}

We employed a Metropolis-within-Gibbs sampler where $\alpha$, $\beta$, $\sigma$, and the elements of $\{\mathbf{r}\}_\text{unobs}$ are updated in turn. We performed updates using random walk Metropolis steps for all parameters with adaptive tuning \citep{haario2001, roberts2009} (also see \citet{craiu2014}) to improve mixing. We ran the adaptive algorithm for $100,000$ iterations, which were subsequently discarded as burn-in, to tune the proposal covariance matrix. We drew the next $100,000$ MCMC samples using the tuned proposal covariance matrix from the first $100,000$ iterations. Finally, we used these $100,000$ iterations for posterior estimation. This procedure took about 90 hours of computational time on a high-performance computing cluster. While not all chains converged in this many iterations, running all chains until convergence would have been infeasible in this framework. Thus, we chose to assess convergence at an individual level and remove the simulated paths which failed to meet the convergence criterion.

\section{Estimation}\label{sec:antest}

The model equation based on \eqref{eq:num030} is
\begin{align}
    g_{x\tau} &= v_{x\tau}\beta + m(\mathbf{r}_\tau)\mathbf{a}_{x\tau}' \boldsymbol{\gamma} + \epsilon_{\tau}\label{eq:ant010}
\end{align}
where 
\begin{align}
    g_{x\tau} &\equiv \frac{x_{\tau+2}-x_{\tau+1}}{h_{\tau+1}} - \frac{x_{\tau+1}-x_{\tau}}{h_{\tau}}\label{eq:ant011}\\
    v_{x\tau} &\equiv x_\tau - x_{\tau + 1}\\
    \boldsymbol{\gamma} &\equiv \begin{bmatrix}
        -\beta p_1 & -\beta p_2 & \hdots & -\beta p_J 
    \end{bmatrix}' \\
    \epsilon_{\tau} &\sim \text{N}(0, h_\tau \sigma^2 m^2(\mathbf{r}_\tau)).\label{eq:ant014}
\end{align} 
Element $u=1, \hdots, J$ of column vector $\mathbf{a}_{x\tau}$ is $\frac{1}{2} h_\tau$ if grid cell $u$ contains $\begin{bmatrix}
    x_{\tau}+1 & y_\tau
\end{bmatrix}'$, $-\frac{1}{2} h_\tau$ if grid cell $u$ contains $\begin{bmatrix}
    x_{\tau}-1 & y_\tau
\end{bmatrix}'$, and 0 otherwise. This formulation comes from the use of a raster-based centered difference equation to approximate the gradient of the potential surface. We include $h_\tau$ in $\mathbf{a}_{x\tau}$ instead of keeping the two separate for ease of vector notation in \eqref{eq:ant010}. If we multiply $\mathbf{a}_{x\tau}'$ by a vector of grid cell values, we get the product of $h_\tau$ and the $x$ component of the estimated gradient of the gridded surface at $\mathbf{r}_\tau$. Thus, we approximate
\begin{align*}
    \beta h_\tau m(\mathbf{r_\tau})\left[-\frac{\partial}{\partial x}p( \mathbf{r}_\tau )\right]
    &\approx m(\mathbf{r_\tau})\mathbf{a}_{x\tau}'\boldsymbol{\gamma}\\
    &= -\beta h_\tau m(\mathbf{r_\tau})\frac{p\left(\begin{bmatrix}
    x_{\tau}+1 & y_\tau
\end{bmatrix}'\right) - p\left(\begin{bmatrix}
    x_{\tau}-1 & y_\tau
\end{bmatrix}'\right)}{2}.
\end{align*}

Similarly, we could construct \eqref{eq:ant010}--\eqref{eq:ant014} in the $y$ direction by replacing each $x$ with $y$. Element $u=1, \hdots, J$ of $\mathbf{a}_{y\tau}$ is $\frac{1}{2} h_\tau$ if grid cell $u$ contains $\begin{bmatrix}
    x_{\tau} & y_\tau+1
\end{bmatrix}'$, $-\frac{1}{2} h_\tau$ if grid cell $u$ contains $\begin{bmatrix}
    x_{\tau} & y_\tau-1
\end{bmatrix}'$, and 0 otherwise.

We estimate $\mathbf{p} \equiv \begin{bmatrix}
    p_1 & \hdots & p_J
\end{bmatrix}'$ and $\mathbf{m} \equiv \begin{bmatrix}
    m_1 & \hdots & m_J
\end{bmatrix}'$ with an iterative procedure. Our proposed approach is as follows:
\begin{enumerate}
    \item Obtain a preliminary estimate of mean parameters ($\beta$ and $\mathbf{p}$) assuming the motility surface is constant (model errors in \eqref{eq:ant014} are independent and identically distributed).
    \item Estimate variance parameters ($\mathbf{m}$) using residuals from step 1.
    \item Estimate mean parameters ($\beta$ and $\mathbf{p}$) conditioned on the variance estimates from step 2.
\end{enumerate}

In step 1, we assume the motility surface is approximately constant (i.e., $m_j$ is similar to the motility surface evaluated in grid cells adjacent to cell $j$) so that the error variance in \eqref{eq:ant014} is constant and we can absorb $m( \mathbf{r}_\tau )$ in the estimate of $\boldsymbol{\gamma}$ when we fit the model \eqref{eq:ant010}. Let $\mathbf{E} \equiv \begin{bmatrix}
    \mathbf{v} & \mathbf{A}
\end{bmatrix}$ where $\mathbf{v}$ is a column vector of all $v_{x\tau}$ and $v_{y\tau}$ that are not in the holdout set and $\mathbf{A}$ is a matrix with rows consisting of all $\mathbf{a}_{x\tau}'$ and $\mathbf{a}_{y\tau}'$ that are not in the holdout set. Let $\mathbf{g}$ be the column vector containing all $g_{x\tau}$ and $g_{y\tau}$ that are not in the holdout set. When combining over the indices $x\tau$ and $y\tau$, we ensure the combined values are ordered in the same way each time by ant, time, and direction. 

We obtain preliminary estimates of $\beta$ and $\boldsymbol{\gamma}$ using penalized least squares estimation
\begin{align*}
    \begin{bmatrix}
        \hat{\beta} \\
        \hat{\boldsymbol{\gamma}}
    \end{bmatrix} &= \text{argmin}_{\beta, \boldsymbol{\gamma}}
    \left\{
    \left( \mathbf{g} - \mathbf{E}\begin{bmatrix}
        \beta \\
        \boldsymbol{\gamma}
    \end{bmatrix} \right)'
    \left( \mathbf{g} - \mathbf{E}\begin{bmatrix}
        \beta \\
        \boldsymbol{\gamma}
    \end{bmatrix} \right)
    + \lambda \begin{bmatrix}
        \beta \\
        \boldsymbol{\gamma}
    \end{bmatrix}'\mathbf{Q}\begin{bmatrix}
        \beta \\
        \boldsymbol{\gamma}
    \end{bmatrix}
    \right\}
    \\
    &= (\mathbf{E}'\mathbf{E}+\lambda \mathbf{Q})^{-1}\mathbf{E}'\mathbf{g}
\end{align*}
where the $(J+1) \times (J+1)$ penalty matrix $\mathbf{Q}$ ensures the potential surface is smooth. 
To bypass penalization in estimation of $\beta$, we let the first row and column of $\mathbf{Q}$ be $\mathbf{0}$ vectors. If we ignore the first row and column of $\mathbf{Q}$, the elements of the $J\times J$ submatrix are $\{\mathbf{Q}_{ij}:i,j=1,\hdots,J\}$. $\mathbf{Q}$ penalizes the sum of squared first differences between the estimated potential surface in neighboring grid cells, i.e.,
\begin{align*}
    \begin{bmatrix}
        \beta \\
        \boldsymbol{\gamma}
    \end{bmatrix}'\mathbf{Q}\begin{bmatrix}
        \beta \\
        \boldsymbol{\gamma}
    \end{bmatrix} &= \sum_{i,j \text{ adjacent grid cells}}
    \left(
    \gamma_i - \gamma_j
    \right)^2 \\
    &= \sum_{i,j \text{ adjacent grid cells}}
    \left(
    \beta p_j - \beta p_i
    \right)^2 .
\end{align*}
This is accomplished by letting $\mathbf{Q}_{ii}$ be the count of grid cells adjacent to grid cell $i$. The off-diagonal elements $\{\mathbf{Q}_{ij}:i\neq j\}$ are defined
\begin{align*}
    \mathbf{Q}_{ij}=\begin{cases}
        -1, & \text{grid cells $i$ and $j$ are adjacent}\\
        0, & \text{otherwise.}
    \end{cases}
\end{align*} 
As the tuning parameter $\lambda$ increases, the estimated potential surface becomes smoother. This is similar to putting a conditional autoregressive prior on $\boldsymbol{\gamma}$.

In step 2, we use the residual term 
\begin{align*}
    \boldsymbol{\hat\epsilon} &\equiv \mathbf{g} - \mathbf{E}\begin{bmatrix}
        \hat{\beta} \\
        \hat{\boldsymbol{\gamma}}
    \end{bmatrix}
\end{align*}
from the preliminary estimation procedure in step 1 to estimate the motility surface. Similar to restricted maximum likelihood estimation, we estimate error variance using $\boldsymbol{\hat\epsilon}^2$. In \eqref{eq:ant014}, $\text{Var}(\epsilon_{\tau}) = \text{E}(\epsilon_{\tau}^2) = h_\tau \sigma^2 m^2(\mathbf{r}_\tau)$. We want to estimate $m^2(\mathbf{r}_\tau) = \text{E}(\epsilon_{\tau}^2h_
\tau^{-1}\sigma^{-2}) = \text{E}(\epsilon_{\tau}^2h_
\tau^{-1})$ (recall $\sigma = 1$). To avoid negative estimates of $m^2(\mathbf{r}_\tau)$, we estimate $\text{log}(m^2(\mathbf{r}_\tau))$ instead. We fit a generalized additive model with mean
\begin{align}
    \text{E}\left[\text{log}(\hat\epsilon_{\tau}^2 h_\tau^{-1})\right] = \nu + f(x_\tau, y_\tau)\label{eq:ant100}
\end{align}
where $f$ is a non-linear smooth function of location and $\nu$ is the intercept term. We fit this generalized additive model using the \pkg{mgcv} R package \citep{mgcvBook}. While alternative methods including maximum likelihood estimation of the residuals could be implemented, the generalized additive model is fast and results in accurate estimation when the data size is large. We exponentiate the fitted values to obtain estimates of $m^2(\mathbf{r}_\tau)$ for all observations.

We now have the components to easily obtain an estimate $\hat{\mathbf{m}} = \begin{bmatrix}
\hat{m}_1 & \hdots & \hat{m}_J
\end{bmatrix}'$ of the vector $\mathbf{m}$ of motility grid cell values. For $j=1,\hdots, J$,
\begin{align*}
    \hat{m}_j &\equiv \hat{\nu} + \hat{f}(\mathbf{c}_{j})
\end{align*}
where $\hat{\nu}$ and $\hat{f}$ are estimates of $\nu$ and $f$ from the generalized additive model fit and $\mathbf{c}_{j}$ contains the coordinates for the center of grid cell $j$.

In step 3, we use the original model equation \eqref{eq:ant010}, but now we replace ${m}(\mathbf{r}_\tau)$ with $\hat{m}(\mathbf{r}_\tau)$, the estimated motility surface evaluated at $\mathbf{r}_\tau$. Then we divide by $\hat{m}(\mathbf{r}_\tau)h_\tau^{\sfrac{1}{2}}$ to produce a model with independent and identically distributed errors. The resulting model equation is
\begin{align}
    \tilde{g}_{x\tau} &= \tilde{v}_{x\tau}\beta +  \tilde{\mathbf{a}}_{x\tau}' {\boldsymbol{\gamma}} + \tilde{\epsilon}_{x\tau} \label{eq:anttilde1}
\end{align}
where 
\begin{align}
    \tilde{g}_{x\tau} &\equiv \frac{{g}_{x\tau}}{\hat{m}(\mathbf{r}_\tau)h_\tau^{\sfrac{1}{2}}}\label{eq:anttilde2}\\
    \tilde{v}_{x\tau} &\equiv \frac{{v}_{x\tau}}{\hat{m}(\mathbf{r}_\tau)h_\tau^{\sfrac{1}{2}}}\label{eq:anttilde3}\\
    \tilde{\mathbf{a}}_{x\tau}' &\equiv h_\tau^{-\sfrac{1}{2}}\mathbf{a}_{x\tau}'\label{eq:anttilde4}\\
    {\boldsymbol{\gamma}} &\equiv \begin{bmatrix}
        -\beta p_1 & -\beta p_2 & \hdots & -\beta p_J 
    \end{bmatrix}' \label{eq:anttilde5}\\
    \tilde{\epsilon}_{x\tau} &\sim \text{N}(0, 1).\label{eq:anttilde6}
\end{align}
As in the original model equation, we construct  \eqref{eq:anttilde1}--\eqref{eq:anttilde6} similarly in the $y$ direction. Let $\tilde{\mathbf{E}} \equiv \begin{bmatrix}
    \tilde{\mathbf{v}} & \tilde{\mathbf{A}}
\end{bmatrix}$ where $\tilde{\mathbf{v}}$ is a column vector of all $\tilde{v}_{x\tau}$ and $\tilde{v}_{y\tau}$ and $\tilde{\mathbf{A}}$ is a matrix with rows of all $\tilde{\mathbf{a}}_{x\tau}'$ and $\tilde{\mathbf{a}}_{y\tau}'$. Let $\tilde{\mathbf{g}}$ be the column vector containing all $\tilde{g}_{x\tau}$ and $\tilde{g}_{y\tau}$. This time, we obtain final estimates of $\beta$ and ${\boldsymbol{\gamma}}$
\begin{align*}
    \begin{bmatrix}
        \hat{\beta}^* \\
        \hat{{\boldsymbol{\gamma}}}^*
    \end{bmatrix} &\equiv
    (\tilde{\mathbf{E}}'\tilde{\mathbf{E}}+\lambda \mathbf{Q})^{-1}\tilde{\mathbf{E}}'\tilde{\mathbf{g}}.
\end{align*}

    Finally, we estimate $\mathbf{p}$, the vector of potential grid cell values, with
\begin{align*}
    \widehat{\mathbf{p}} &\equiv -\frac{\hat{{\boldsymbol{\gamma}}}^*}{\hat{\beta}^*}.
\end{align*}
As explained in step 1, the penalty matrix $\mathbf{Q}$ ensures the estimated potential surface is smooth. Since the potential surface only enters the model through its gradient, the potential surface is only identifiable up to an unknown additive constant.

We repeated this 3-step procedure for $\text{log}(\lambda) = -8, -7, \hdots, 7, 8$. We chose the $\text{log}(\lambda)$ that resulted in the lowest mean squared prediction error
\begin{align*}
    \left[\mathbf{g}_\text{holdout} - \hat{\mathbf{g}}_\text{holdout}\right]'\left[\mathbf{g}_\text{holdout} - \hat{\mathbf{g}}_\text{holdout}\right],
\end{align*}
where
\begin{align*}
    \hat{\mathbf{g}}_\text{holdout} &\equiv 
    \begin{bmatrix}
        \mathbf{v}_\text{holdout} & 
        \text{diag}[\hat{m}(\mathbf{r}_{ \text{holdout}})]
        \mathbf{A}_\text{holdout}
    \end{bmatrix}
    \begin{bmatrix}
        \hat{\beta}^* \\
        \hat{{\boldsymbol{\gamma}}}^*
    \end{bmatrix}
\end{align*}
and $\mathbf{g}_\text{holdout}$, $\mathbf{v}_\text{holdout}$, and $\mathbf{A}_\text{holdout}$ are identical to $\mathbf{g}$, $\mathbf{v}$, and $\mathbf{A}$ in step 1 except now they consist of only the holdout data. $\hat{m}(\mathbf{r}_{ \text{holdout}})$ is a vector of estimated motility values at the positions in the holdout data.

\section{Simulation study of the estimation procedure}

As outlined in Supplementary Materials \ref{sec:antest}, step 2 of the estimation procedure entails estimation of the squared motility surface evaluated at each observation, i.e., $m^2(\textbf{r}_\tau)=\text{E}(\epsilon_\tau^2 h_\tau^{-1})$. We achieve this by estimating $\text{E}[\text{log}(m^2(\textbf{r}_\tau))]$ and exponentiating the fitted values. We recognize that this procedure may introduce bias, and thus we have conducted a simulation study to quantify this bias. We chose to simulate movement from simpler potential and motility surfaces than those fit with the ant data to avoid dealing with nest boundaries. Instead, we constructed a large surface of which only the center is utilized and analyzed our results in the areas where simulated data is present.

The data were simulated from an SDE model \eqref{eq:sde010}-\eqref{eq:sde020} with a linear motility and quadratic potential surface
\begin{align*}
m(x_\tau, y_\tau)&\equiv 0.02y_\tau + 2\\
p(\mathbf{r}_\tau)&\equiv 0.02 (\mathbf{r}_\tau-50)'(\mathbf{r}_\tau-50).
\end{align*}
The numerical approximation \eqref{eq:num030} becomes
\begin{align*}
    {\frac{\mathbf{r}_{\tau+2}-\mathbf{r}_{\tau+1}}{h_{\tau+1}}-\frac{\mathbf{r}_{\tau+1}-\mathbf{r}_{\tau}}{h_\tau}}&=\beta h_\tau \left([0.02y_\tau+2][-0.04 (\mathbf{r}_\tau - 50)]-\frac{\textbf{r}_{\tau+1}-\mathbf{r}_\tau}{h_\tau}\right)\\&\ \ +h_\tau^{\sfrac{1}{2}} (0.02y_\tau + 2) \boldsymbol{\epsilon}_\tau
\end{align*}
which, after setting $h_\tau = 1$, we rearrange to produce the simulation generation equation
\begin{align*}
    \mathbf{r}_{\tau+2}&=2\mathbf{r}_{\tau+1} - \mathbf{r}_{\tau} + \beta\left[ (0.02y_\tau+2)(-0.04 (\mathbf{r}_\tau - 50)) - (\mathbf{r}_{\tau+1}-\mathbf{r}_{\tau}) \right] + (0.02y_\tau + 2) \boldsymbol{\epsilon}_\tau.
\end{align*}

\begin{figure}[h]
\centering\includegraphics[width=.9\linewidth]{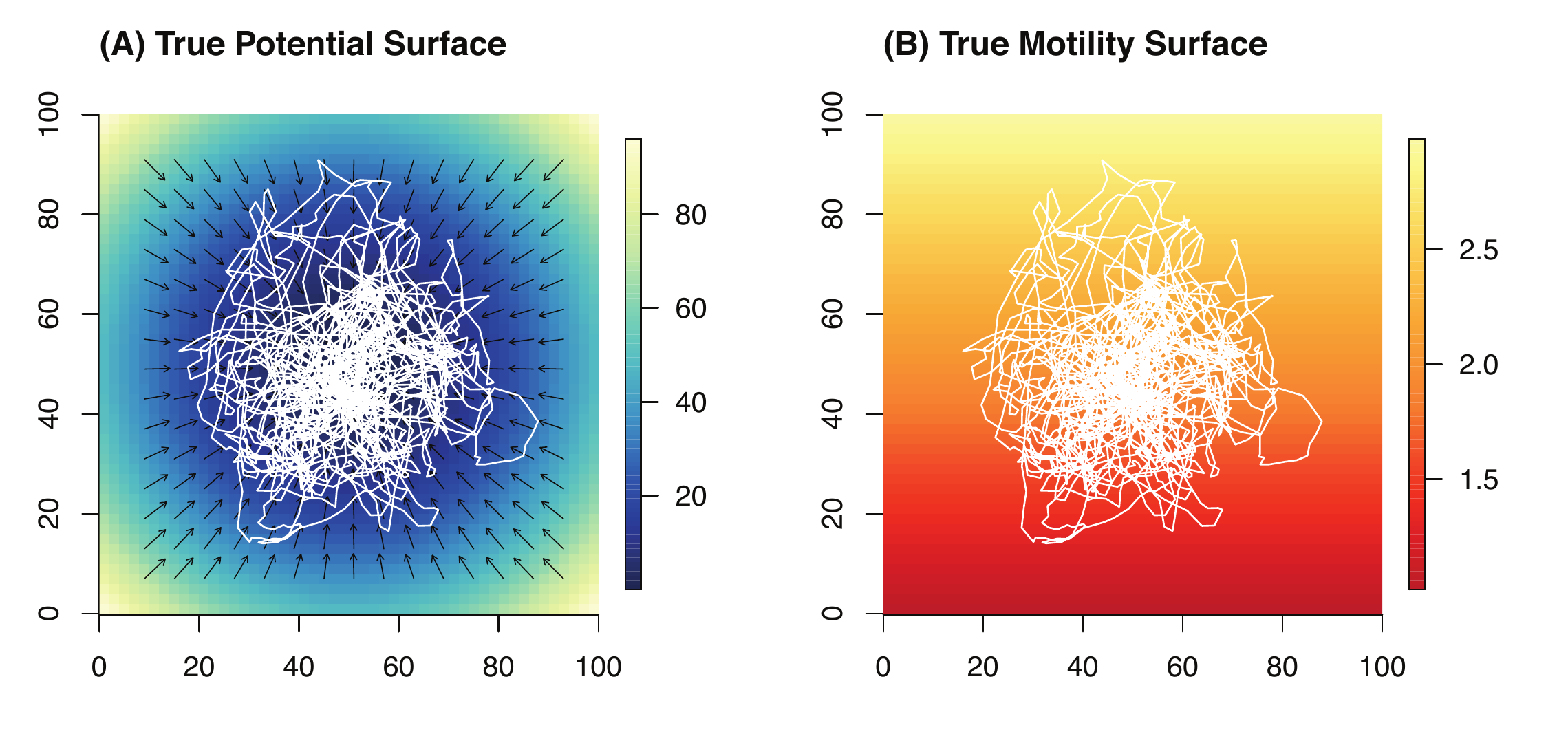}
\caption{True potential and motility surfaces with one example simulation shown in white. Black gradient vectors on the potential surface depict the negative gradient scaled by 5.}
\label{fig:simstudy010}
\end{figure}

We performed 100 simulation runs, each of which generated 5 individual paths for 2000 time steps. For illustrative purposes, we plotted the simulated path of one individual on the potential and motility surfaces which generated the simulations in Figure \ref{fig:simstudy010}. We fit each of these simulations using the estimation procedure in Section \ref{sec:est} to produce 100 pairs of estimated motility and potential surfaces. Each surface is a $50\times 50$ grid of values, where each grid cell is $2\times 2$. When comparing surfaces, we subset the surfaces to where the majority of the data were located: within a radius of 23 from the center $(50,50)$. On average, $77.95\%$ of the simulated observations fell within the radius of 23 (standard deviation $= 1.21\%$).

\begin{figure}[h]
\centering\includegraphics[width=\linewidth]{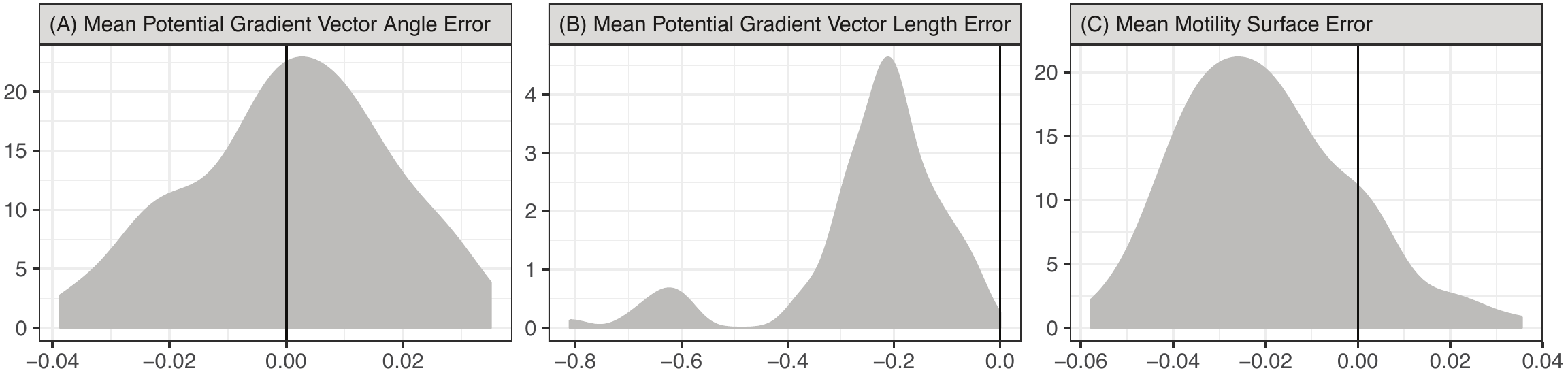}
\caption{Density estimates for errors in the model fit on simulated data. The black vertical line is positioned at 0.}
\label{fig:simstudy020}
\end{figure}

\begin{figure}[h]
\centering\includegraphics[width=.7\linewidth]{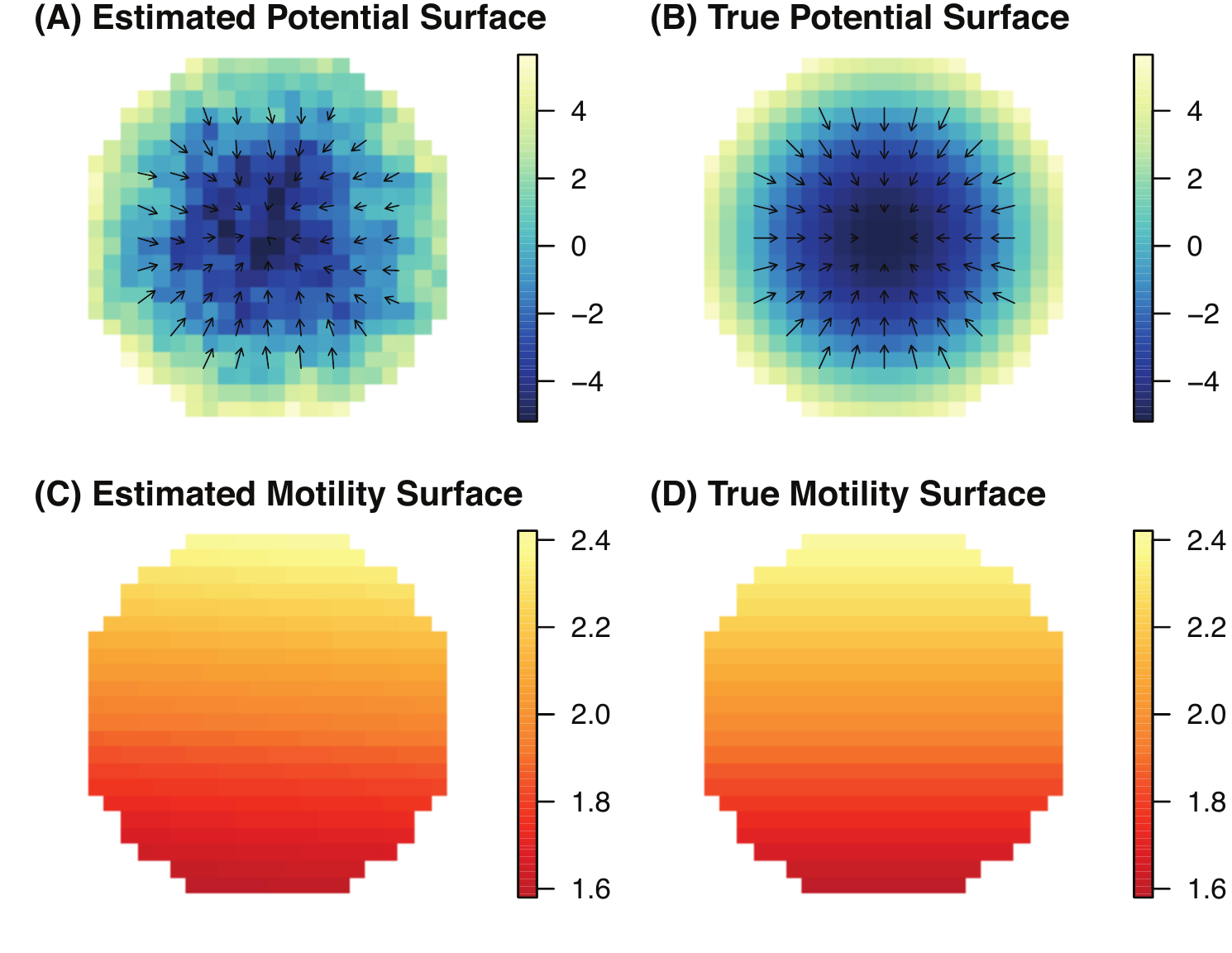}
\caption{For one randomly selected simulation, the estimated potential (A) and motility (B) surfaces are compared to the true potential (B) and motility (D) surfaces.}
\label{fig:simstudy030}
\end{figure}

We assessed motility and potential surface estimation using three metrics: mean potential gradient vector angle, the average angle between the gradient vectors for the estimated and true potential surfaces; mean potential gradient vector length error, the average of the magnitudes of the estimated potential surface gradient vector minus those of the true potential surface gradient vector, and mean motility surface error, the average of the grid cell values of the estimated motility surface minus those of the true motility surface. The empirical density of the metrics for the 100 simulations are displayed in Figure \ref{fig:simstudy020}. As shown in Figure \ref{fig:simstudy020}(A), the average angle of the potential surface gradient vectors is unbiased, as we would expect. As in \ref{fig:simstudy020}(B), the average error in the magnitude of the potential surface gradient vectors is always negative, which is also unsurprising since the estimation procedure involves smoothing. In \ref{fig:simstudy020}(C), we see that the motility surface is estimated well in simulations. Although the motility surface is underestimated on average in 84 of the 100 simulations, the interpretation of the results is not impacted by this slight underestimation. In Figure \ref{fig:simstudy030}, we randomly chose one simulation to display next to the true potential and motility surfaces. The potential surfaces are centered at zero since they are unidentifiable up to an additive constant.

\FloatBarrier
\FloatBarrier

\FloatBarrier
\FloatBarrier

\FloatBarrier
\FloatBarrier

\end{document}